\documentclass{tlp}

\usepackage{aopmath}
\usepackage{latexsym}
\usepackage{amssymb}

\newtheorem{definition}{Definition} 
\newtheorem{example}{Example} 

\long\def\comment#1{}

\newcommand{\foo}[1]{#1}
\newcommand{\toppos}{\mbox{\footnotesize$\Lambda$}} 
\newcommand{\nat}{\mbox{$I\!\!N$}}
\newcommand{\nil}{[\:]}
\newcommand{\ol}[1]{\overline{#1}}  
\newcommand{\sleq}{\leqslant}
\newcommand{\pr}[1]{\mbox{\tt #1}}   
\newcommand{\Control}{\foo{Expr}}
\newcommand{\caseof}[1]{\foo{case}\;#1\;\foo{of}\;}
\newcommand{\fcaseof}[1]{\foo{fcase}\;#1\;\foo{of}\;}
\newcommand{\To}{\Longrightarrow}
\newcommand{\Var}{{\cal V}ar} 

\let\l=\langle
\let\r=\rangle
\def \tuple#1{\langle #1 \rangle}

\newcommand{\sql}{\mbox{$[\![$}}
\newcommand{\sqr}{\mbox{$]\!]$}}

\def\defemb#1#2{\expandafter\def\csname #1\endcsname
                              {\relax\ifmmode #2\else\hbox{$#2$}\fi}}
\defemb{cC}{{\cal C}} 
\defemb{cE}{{\cal E}} 
\defemb{cF}{{\cal F}} 
\defemb{cM}{{\cal M}} 
\defemb{cR}{{\cal R}} 
\defemb{cS}{{\cal S}} 
\defemb{cT}{{\cal T}} 
\defemb{cV}{{\cal V}} 
\defemb{cX}{{\cal X}} 

\makeatletter
\newenvironment{prog}{\vspace{1.0ex}\par
\setlength{\parindent}{2ex}
\setlength{\parskip}{0.0ex}
\obeylines\@vobeyspaces\tt}{\vspace{1.0ex}\noindent
}
\makeatother
\newcommand{\startprog}{\begin{prog}}
\newcommand{\stopprog}{\end{prog}\noindent}

\begin{document}
\bibliographystyle{acmtrans}

\title[Forward Slicing by Partial Evaluation]
{Forward Slicing of Functional Logic Programs\\ by Partial Evaluation\thanks{%
    A preliminary short version of this paper appeared in the
    Proceedings of the \emph{12th International Workshop on Logic
      Based Program
      Synthesis and Transformation} (LOPSTR 2002).\protect\\
    This work has been partially supported by the EU (FEDER) and the
    Spanish MEC under grants TIN2004-00231 and TIN2005-09207-C03-02,
    and by the ICT for EU-India Cross-Cultural Dissemination Project
    ALA/95/23/2003/077-054.  }}

\author[Josep Silva and Germ\'an Vidal]
{
JOSEP SILVA and GERM\'AN VIDAL\\
DSIC, Technical University of Valencia\\ 
Camino de Vera s/n, E-46022 Valencia, Spain\\ 
\email{$\{$jsilva,gvidal$\}$@dsic.upv.es}
}

\submitted{5 January 2004}
\revised{20 February  2005}
\accepted{5 January 2006}

\maketitle

\begin{abstract}
  Program slicing has been mainly studied in the context of imperative
  languages, where it has been applied to a wide variety of software
  engineering tasks, like program understanding, maintenance,
  debugging, testing, code reuse, etc. This work introduces the first
  forward slicing technique for declarative multi-paradigm programs
  which integrate features from functional and logic programming.
  Basically, given a program and a \emph{slicing criterion} (a
  function call in our setting), the computed forward slice contains
  those parts of the original program which are \emph{reachable} from
  the slicing criterion. Our approach to program slicing is based on
  an extension of (online) partial evaluation. Therefore, it provides
  a simple way to develop program slicing tools from existing partial
  evaluators and helps to clarify the relation between both
  methodologies. A slicing tool for the multi-paradigm language Curry,
  which demonstrates the usefulness of our approach, has been
  implemented in Curry itself.
\end{abstract}


\section{Introduction}

Essentially, program slicing is a method for decomposing programs by
analyzing their data and control flow. It was first proposed as a
debugging tool to allow a better understanding of the portion of code
which revealed an error. Since this concept was originally introduced
by Weiser \citeyear{Wei79,Wei84}|in the context of imperative
programs|it has been successfully applied to a wide variety of
software engineering tasks (e.g., program understanding, maintenance,
debugging, merging, testing, code reuse).  Surprisingly, there are
very few approaches to program slicing in the context of
\emph{declarative} programming (see Section~\ref{related}).

Roughly speaking, a \emph{program slice} consists of those program
statements which are (potentially) related with the values computed at
some program point and/or variable, often given by a pair
\texttt{(line number, variable)}, referred to as a \emph{slicing
  criterion}.  Program slices are usually computed from a
\emph{program dependence graph} \cite{FOW87,KKPLW81} that makes
explicit both the data and control dependences for each operation in a
program. Program dependences can be traversed backwards and
forwards|from the slicing criterion|giving rise to so-called
\emph{backward} and \emph{forward} slicing, respectively.

\begin{figure}
\fbox{
\begin{minipage}{91ex}
\begin{tabular}{lll}
\tt  (1) read(n); & \tt  (1) read(n); & \tt (1)  \\
\tt  (2) i := 1;  & \tt (2) i := 1; & \tt (2)  \\
\tt  (3) sum := 0; & \tt (3)           & \tt (3) sum := 0; \\
\tt  (4) prod := 1; & \tt (4) prod := 1;  & \tt (4) \\
\tt  (5) while i <= n do & \tt (5) while i <= n do & \tt (5)  \\
\tt  (6)\hspace{3ex} sum := sum + i; & \tt (6) & \tt (6)\hspace{3ex} sum := sum + i; \\
\tt  (7)\hspace{3ex} prod := prod * i; 
            & \tt (7)\hspace{3ex} prod := prod * i; & \tt (7)\\
\tt   (8)\hspace{3ex} i := i + 1; & \tt (8)\hspace{3ex} i := i + 1; 
            & \tt (8) \\
\tt   (9) write(sum); & \tt (9) & \tt (9) write(sum); \\
\tt   (10)write(prod); & \tt (10)write(prod); & \tt (10) \\[2ex]
\hspace{8ex}(a) & \hspace{8ex}(b) & \hspace{8ex}(c) \\
\end{tabular}
\end{minipage}}
\caption{Forward and backward slicing ---an example} \label{fbslice}
\end{figure}  

Essentially, a backward slice consists of the parts of the program
that (potentially) affect the values computed at the slicing
criterion. In contrast, a forward slice consists of the statements
which are dependent on the slicing criterion, a statement being
\emph{dependent} on the slicing criterion if the values computed at
that statement depend on the values computed at the slicing criterion
or if the values computed at the slicing criterion determine if the
statement under consideration is executed \cite{Tip95}.  Consider,
e.g., the example \cite{Tip95} depicted in Fig.~\ref{fbslice}~(a) for
computing the sum and the product of the sequence of numbers
\texttt{1,2,\ldots,n}. Fig.~\ref{fbslice}~(b) shows a backward slice
of the program w.r.t.\ the slicing criterion \texttt{(10,prod)} while
Fig.~\ref{fbslice}~(c) shows a forward slice w.r.t.\ the slicing
criterion \texttt{(3,sum)}.

Additionally, slices can be \emph{dynamic} or \emph{static}, depending
on whether a concrete program's input is provided or not. \emph{Quasi}
static slicing was the first attempt to define a hybrid method ranging
between static and dynamic slicing \cite{Ven91}. It becomes useful
when only the value of some parameters is known. This notion is
closely related to partial evaluation \cite{JGS93}, a well-known
technique to specialize programs w.r.t.\ part of their input data.
For instance, quasi static slicing has been applied to program
understanding by Harman et al.\ \citeyear{HDS95}; similarly, Blazy and
Facon \citeyear{BF98} use partial evaluation for the same purpose.

All approaches to slicing mentioned so far are \emph{syntax
  preserving}, i.e., they are mainly obtained from the original
program by statement deletion. In contrast, \emph{amorphous} slicing
\cite{HD97} exploits different program transformations in order to
simplify the program while preserving its semantics w.r.t.\ the
slicing criterion.  From this perspective, partial evaluation could
straightforwardly be seen as an amorphous slicing technique.
More detailed information on program slicing can be found in the
surveys of Harman and Hierons \citeyear{HH01} and Tip
\citeyear{Tip95}.

The aim of this work is the definition of a forward slicing technique
for a multi-paradigm declarative language which integrates features
from functional and logic programming, like, e.g., Curry
\cite{CurryTR} or Toy \cite{LS99}. Similarly to in \cite{RT96}, where
a first-order functional language is considered, given a program $p$
and a projection function $\pi$, backward slicing should extract a
program that behaves like $\pi(p)$ (e.g., by symbolically pushing
$\pi$ backwards through the body of $p$). For instance, it can be used
to extract a program slice for computing the number of lines in a
string from a more general program that returns a tuple with both the
number of lines and the number of characters of the string; this is
the example that illustrates the backward slicing technique of Reps
and Turnidge \citeyear{RT96}. Such a slicing technique is considered
\emph{backward} because the algorithm proceeds from (part of) the
result backwards to the initial function call, i.e., in the inverse
direction of the standard operational semantics. In contrast, here we
consider the definition of a forward slicing technique that, given a
program and a function call, extracts a program containing all the
statements which are \emph{reachable} from the slicing criterion. Our
slicing technique is considered \emph{forward} because it proceeds
from a given function call to its result, i.e., we follow the control
flow of the standard operational semantics.

Furthermore, rather than defining a new technique from scratch, we
exploit the similarities between slicing and \emph{partial evaluation}
\cite{JGS93}. Since a partial evaluator for the considered language
already exists, our approach provides a simple way to develop a
program slicing tool. The main purpose of partial evaluation is to
specialize a program w.r.t.\ part of its input data and, hence, it is
also known as \emph{program specialization}.  The partially evaluated
program will be (hopefully) executed more efficiently since those
computations that depend only on the known data are performed|at
partial evaluation time|once and for all.  Many (\emph{online})
partial evaluation schemes follow a common pattern: given a program
and a function call (possibly containing partial data structures by
means of free variables), the partial evaluator builds a \emph{finite}
representation|generally a graph|of the possible executions of the
initial call and, then, systematically extracts a \emph{residual}
program|the partially evaluated program|from this graph.

The essence of our approach can be summarized as follows. First, we
consider that, in our functional logic context, a function
call|possibly containing free variables|may also play the role of
\emph{slicing criterion}. Since such a call may have an infinite
computation space, a primary task of both slicing and partial
evaluation is the construction of a finite representation of its
possible program executions. Here, the same algorithm which is used in
partial evaluation can be applied for computing this finite
representation, which will be later used to identify the program
statements that are reachable from the slicing criterion. Then, we
only need to replace the construction of a residual program in partial
evaluation by a simpler post-processing stage that extracts an
executable program which includes the reachable program statements.

While partial evaluation usually achieves its effects by compressing
paths in the graph and by renaming expressions in order to remove
unnecessary function symbols,
slicing should preserve the structure of the original program (here,
we do not consider amorphous slicing): statements can be|totally or
partially|deleted but no new statements can be introduced. In order to
further clarify the relation between partial evaluation and slicing,
let us recall the following classification of partial evaluators
introduced by Gl\"uck and S{\o}rensen \citeyear{GS96}.  According to
this classification, a partial evaluator is
\begin{itemize}
\item \emph{Monovariant}: if each function of the original program
  gives rise to (at most) one residual function;
\item \emph{Polyvariant}: if each function of the original program may
  give rise to one or more residual functions;
\item \emph{Monogenetic}: if each residual function stems from one
  function of the original program;
\item \emph{Polygenetic}: if each residual function may stem from one
  or more functions of the original program.
\end{itemize}
The main contribution of this work is to demonstrate that a forward
slicing technique for functional logic programs can be obtained by
slightly extending a \emph{monovariant} and \emph{monogenetic} partial
evaluation scheme. 
Unfortunately, this kind of monovariant/monogenetic partial evaluation
could be rather imprecise, thus resulting in unnecessarily large
residual programs (i.e., slices). In order to overcome this drawback,
we consider the definition of an extended operational semantics to
perform partial evaluations, which helps us to preserve as much
information as possible while maintaining the monovariant/monogenetic
nature of the process.

The main contributions of this work can be summarized as follows:
\begin{itemize}
\item We define the first forward slicing technique for functional
  logic programs. Furthermore, the application of our developments to
  (first-order) lazy functional programs would be straightforward,
  since either the syntax and the underlying (online) partial
  evaluators---e.g., positive supercompilation \cite{SGJ93}---share
  many similarities.
  
\item We do not need to consider separately static and dynamic
  slicing, since the underlying partial evaluation scheme naturally
  accepts partial input data.
  
\item Our method is defined in terms of an existing partial evaluation
  scheme and, thus, it is easy to implement by adapting current
  partial evaluators.
  
\item Finally, our approach helps to clarify the relation between
  forward slicing and (online) partial evaluation.
\end{itemize}
This paper is organized as follows. In the next section we recall some
foundations for understanding the subsequent developments.
Section~\ref{pe} introduces a notion of forward slicing in the context
of functional logic programming. We then recall, in Section~\ref{pe1},
the narrowing-driven approach to partial evaluation.
Section~\ref{depen} defines an algorithm for computing program
dependences by partial evaluation, while Section~\ref{extract} uses
these dependences to extract program slices.
Section~\ref{exp} presents a prototype implementation of the program
slicing tool and show some selected experiments.  Several related
works are discussed in Section~\ref{related} before we conclude in
Section~\ref{sec-concl}.  Proofs of technical results can be found in
\ref{proofs}.

\section{Foundations}
\label{sec-foundations}

We recall in this section some basic notions of term rewriting
\cite{BN98,Klo91,Ter03} and functional logic programming
\cite{Han94JLP}.

\subsection{Preliminaries}

Throughout this paper, we consider a (\emph{many-sorted})
\emph{signature} $\Sigma$ partitioned into a set $\cC$ of
\emph{constructors} and a set $\cF$ of (defined) \emph{functions} or
\emph{operations}.  We write $c/n \in \cC$ and $f/n \in \cF$ for
$n$-ary constructor and operation symbols, respectively.  There is at
least one sort $Bool$ containing the constructors \pr{True} and
\pr{False}.  The set of \emph{terms} and \emph{constructor terms} with
\emph{variables} (e.g., $x,y,z$) from $\cX$ are denoted by $\cT(\cC
\cup \cF,\cX)$ and $\cT(\cC,\cX)$, respectively. A term is
\emph{linear} if it does not contain multiple occurrences of one
variable. The set of variables occurring in a term $t$ is denoted by
$\Var(t)$.  A term $t$ is \emph{ground} if $\Var(t) = \emptyset$.

A \emph{pattern} is a term of the form $f(d_{1},\ldots,d_{n})$ where
$f/n \in \cF$ and $d_1,\ldots,d_n \in \cT(\cC,\cX)$.  A term is
\emph{operation-rooted} (constructor-rooted) if it has an operation
(constructor) symbol at the root.  A \emph{position} $p$ in a term $t$
is represented by a sequence of natural numbers ($\toppos$ denotes the
empty sequence, i.e., the root position).  $t|_p$ denotes the
\emph{subterm} of $t$ at position $p$, and $t[s]_p$ denotes the result
of \emph{replacing the subterm} $t|_p$ by the term $s$.  We denote a
\emph{substitution} $\sigma$ by $\{x_1 \mapsto t_1,\ldots, x_n \mapsto
t_n\}$ where $\sigma(x_i) = t_i$ for $i=1,\ldots,n$ (with $x_{i}\neq
x_{j}$ if $i\neq j$), and $\sigma(x) = x$ for all other variables $x$.
A substitution $\sigma$ is \emph{constructor}, if $\sigma(x)$ is a
constructor term for all $x$.  The identity substitution is denoted by
$id$.  A substitution $\theta$ is more general than $\sigma$, in
symbols $\theta \leq \sigma$, iff there exists a substitution $\gamma$
such that $\gamma\circ\theta = \sigma$ (``$\circ$'' denotes the
composition operator).  Term $t'$ is a (constructor) \emph{instance}
of term $t$ if there is a (constructor) substitution $\sigma$ with $t'
= \sigma(t)$.

A set of rewrite rules (or oriented equations) $l = r$ such that $l
\not\in \cX$, and $\Var(r)\subseteq\Var(l)$ is called a \emph{term
  rewriting system} (TRS). Terms $l$ and $r$ are called the
\emph{left-hand side} and the \emph{right-hand side} of the rule,
respectively.  A TRS $\cR$ is \emph{left-linear} if $l$ is linear for
all $l = r\in\cR$.  A TRS is \emph{constructor-based} if each
left-hand side $l$ is a pattern.  In the following, a functional logic
\emph{program} is a left-linear constructor-based TRS.
A \emph{rewrite step} is an application of a rewrite rule to a term,
i.e., $t \to_{p,R} s$ if there exists a position $p$ in $t$, a rewrite
rule $R= (l = r)$ and a substitution $\sigma$ with $t|_p = \sigma(l)$
and $s = t[\sigma(r)]_p$.  The instantiated left-hand side $\sigma(l)$
of a rule $l = r$ is called a \emph{redex} (\emph{red}ucible
\emph{ex}pression).  Given a relation $\to$, we denote by $\to^\ast$
its transitive and reflexive closure.

\begin{example} \label{ex1}
  Consider the following TRS that defines the addition on natural
  numbers represented by terms built from \pr{Zero} and
  \pr{Succ}:\footnote{In the examples, we write constructor symbols
    starting with upper case (except for the list constructors,
    ``\pr{$\nil$}'' and ``\pr{:}'', which are a shorthand for \pr{Nil}
    and \pr{Cons}, respectively).}
  \[
  \begin{array}{r@{~~=~~}l@{~~~~}l}
    \tt Zero + y & \tt y & \tt (R_1) \\
    \tt Succ(x) + y & \tt Succ(x+y) & \tt (R_2) \\
  \end{array}
  \]
  Given the term $\tt Succ(Zero)+Succ(Zero)$, we have the following
  sequence of rewrite steps:
  \[
  \begin{array}{lll}
  \tt Succ(Zero)+Succ(Zero) & \tt \to_{\toppos,R_2} & \tt Succ(Zero + Succ(Zero) \\
  & \tt \to_{1,R_1} & \tt Succ(Succ(Zero))
  \end{array}
  \]
\end{example}

\subsection{Narrowing}

Functional \emph{logic} programs mainly differ from purely functional
programs in that function calls may contain \emph{free} variables.  In
order to evaluate terms containing free variables, \emph{narrowing}
non-deterministically instantiates these variables so that a rewrite
step is possible. Formally, $t \leadsto_{(p,R,\sigma)} t'$ is a
\emph{narrowing step} if $p$ is a non-variable position of $t$ and
$\sigma(t) \to_{p,R} t'$.  We often write $t \leadsto_{\sigma} t'$
when the position and the rule are clear from the context.  We denote
by $t_0 \leadsto^{n}_\sigma t_n$ a sequence of $n$ narrowing steps
$t_0 \leadsto_{\sigma_1} \ldots \leadsto_{\sigma_n} t_n$ with $\sigma
= \sigma_n \circ \cdots \circ \sigma_1$ (if $n = 0$ then $\sigma =
id$), usually restricted to the variables of $t_0$.
Due to the presence of free variables, a term may be reduced to
different values after instantiating these variables to different
terms.  Given a narrowing derivation $t_0 \leadsto^{\ast}_\sigma t_n$,
we say that $t_{n}$ is a computed \emph{value} and $\sigma$ is a
computed \emph{answer} for $t_{0}$.

\begin{example}
  Consider again the definition of function ``\pr{+}'' in
  Example~\ref{ex1}. Given the term $\tt x+Succ(Zero)$, narrowing
  non-deterministically performs the following derivations:
  \[
  \begin{array}{lll}
    \tt x + Succ(Zero) & \tt \leadsto_{\toppos,R_1,\{x \mapsto Zero\}} 
            & \tt Succ(Zero) \\[1ex]
    \tt x + Succ(Zero) & \tt \leadsto_{\toppos,R_2,\{x \mapsto Succ(y_1)\}} 
            & \tt Succ(y_1 + Succ(Zero)) \\
            & \tt \leadsto_{1,R_1,\{y_1 \mapsto Zero \}} & \tt Succ(Succ(Zero))\\[1ex]
    \tt x + Succ(Zero) & \tt \leadsto_{\toppos,R_2,\{x \mapsto Succ(y_1)\}} 
            & \tt Succ(y_1 + Succ(Zero)) \\
            & \tt \leadsto_{1,R_2,\{y_1 \mapsto Succ(y_2) \}} 
            & \tt Succ(Succ(y_2 + Succ(Zero)))\\
            & \tt \leadsto_{1.1,R_1,\{y_2 \mapsto Zero\}} 
            & \tt Succ(Succ(Succ(Zero))) \\
    \tt \ldots
  \end{array}
  \]
  Therefore, $\tt x + Succ(Zero)$ non-deterministically computes the
  values 
  \begin{itemize}
  \item $\tt Succ(Zero)$ with answer $\tt \{x \mapsto Zero\}$,
  \item $\tt Succ(Succ(Zero))$ with answer $\tt \{ x \mapsto
    Succ(Zero)\}$,
  \item $\tt Succ(Succ(Succ(Zero)))$ with answer $\tt \{ x \mapsto
    Succ(Succ(Succ(Zero)))\}$, etc.
  \end{itemize}
\end{example}
As in logic programming, narrowing derivations can be represented by a
(possibly infinite) finitely branching \emph{tree}.  Formally, given a
program $\cR$ and an operation-rooted term $t$, a \emph{narrowing
  tree} for $t$ in $\cR$ is a tree satisfying the following
conditions: (a) each node of the tree is a term, (b) the root node is
$t$, (c) if $s$ is a node of the tree then, for each narrowing step $s
\leadsto_{p,R,\sigma} s'$, the node has a child $s'$ and the
corresponding arc is labeled with $(p,R,\sigma)$, and (d) nodes which
are constructor terms have no children.

In order to avoid unnecessary computations and to deal with infinite
data structures, demand-driven generation of the search space has been
advocated by a number \emph{lazy} narrowing strategies
\cite{GLMP91,LLR93,MR92}. Due to its optimality properties w.r.t.\ the
length of derivations and the number of computed solutions,
\emph{needed narrowing} \cite{AEH00} is currently the best lazy
narrowing strategy.

\subsection{Needed Narrowing}

Needed narrowing \cite{AEH00} is defined on \emph{inductively
  sequential} TRSs \cite{Ant92}, a subclass of left-linear
constructor-based TRSs. Essentially, a TRS is \emph{inductively
  sequential} when all its operations are defined by rewrite rules
that, recursively, make on their arguments a case distinction
analogous to a data type (or structural) induction. Inductive
sequentiality is not a limiting condition for programming.  In fact,
the first-order components of many functional (logic) programs written
in, e.g., Haskell, ML or Curry, are inductively sequential.

We say that $s \leadsto_{p,R,\sigma} t$ is a \emph{needed narrowing
  step} iff $\sigma(s) \to_{p,R} t$ is a \emph{needed rewrite} step in
the sense of Huet and L\'evy \citeyear{HL92}, i.e., in every
computation from $\sigma(s)$ to a normal form, either $\sigma(s)|_p$
or one of its \emph{descendants} must be reduced. Here, we are
interested in a particular needed narrowing strategy, denoted by
$\lambda$ in \cite[Def.\ 13]{AEH00} which is based on the notion of a
\emph{definitional tree} \cite{Ant92}, a hierarchical structure
containing the rules of a function definition, which is used to guide
the needed narrowing steps.  This strategy is basically equivalent to
\emph{lazy narrowing} \cite{MR92} where narrowing steps are applied to
the outermost function, if possible, and inner functions are only
narrowed if their evaluation is \emph{demanded} by a constructor
symbol in the left-hand side of some rule (i.e., a typical outermost
strategy). 

\begin{example} \label{ex-leq}
  Consider following rules which define the less-or-equal function on
  natural numbers:
  \[
  \begin{array}{r@{~\sleq~}l@{~~=~~}l}
  \tt Zero    & \tt y & \tt True \\
  \tt Succ(x) & \tt Zero & \tt False \\
  \tt Succ(x) & \tt Succ(y) & \tt x \;\sleq\;  y
  \end{array}
  \]
  In a term like $t_1 \sleq t_2$, it is always necessary to evaluate
  $t_1$ to some \emph{head normal form} (i.e., a variable or a
  constructor-rooted term) since all three rules defining ``$\sleq$''
  have a non-variable first argument. On the other hand, the
  evaluation of $t_2$ is only \emph{needed} if $t_1$ is of the form
  \pr{Succ($t$)}.  Thus, if $t_1$ is a free variable, needed narrowing
  instantiates it to a constructor, here \pr{Zero} or \pr{Succ(x)}.
  Depending on this instantiation, either the first rule is applied or
  the second argument $t_2$ is evaluated.
\end{example}

\subsection{Declarative Multi-Paradigm Languages}

Functional logic languages have recently evolved to so called
declarative \emph{multi-paradigm} languages like, e.g., Curry
\cite{CurryTR}, Toy \cite{HU01} and Escher \cite{Llo94}.  In order to
make things concrete, we consider in this work the language Curry, a
modern multi-paradigm language which integrates features from logic
programming (partial data structures, built-in search), functional
programming (higher-order functions, demand-driven evaluation) and
concurrent programming (concurrent evaluation of constraints with
synchronization on logical variables). Curry follows a Haskell-like
syntax, i.e., variables and function names start with lowercase
letters and data constructors start with an uppercase letter.  The
application of function $f$ to an argument $e$ is denoted by
juxtaposition, i.e., $(f\;e)$.

The basic operational semantics of Curry is based on a combination of
needed narrowing and residuation \cite{Han97b}.  The
\emph{residuation} principle is based on the idea of delaying function
calls until they are ready for a deterministic evaluation.
Residuation preserves the deterministic nature of functions and
naturally supports concurrent computations.  The precise
mechanism---narrowing or residuation---for each function is specified
by \emph{evaluation annotations}. The annotation of a function as
\emph{rigid} forces the delayed evaluation by rewriting, while
functions annotated as \emph{flexible} can be evaluated in a
non-deterministic manner by narrowing.

In actual implementations, e.g., the PAKCS environment \cite{PAKCS00}
for Curry, programs may also include a number of additional features:
calls to external (built-in) functions, concurrent constraints,
higher-order functions, overlapping left-hand sides, guarded
expressions, etc. In order to ease the compilation of programs as well
as to provide a common interface for connecting different tools
working on source programs, a \emph{flat representation} for programs
has recently been introduced.  This representation is based on the
formulation of Hanus and Prehofer \citeyear{HP99} to express
pattern-matching by case expressions.  The complete flat
representation is called FlatCurry \cite{PAKCS00} and is used as an
intermediate language during the compilation of source programs.

In order to simplify the presentation, we will only consider the
\emph{core} of the flat representation. Extending the developments in
this work to the remaining features is not difficult and, indeed, the
implementation reported in Section~\ref{exp} covers many of these
features. The syntax of flat programs is summarized in
Fig.~\ref{flat}, where $\ol{o_n}$ stands for the \emph{sequence} of
objects $o_1,\ldots,o_n$.
\begin{figure}[t]
\rule{\linewidth}{.01in}
\[
\hspace{-5ex}
\begin{array}{lcl@{~~}l@{~~~~}lcl@{~~}l}
\cR & ::= & D_1 \ldots D_m & \mbox{ (program) } 
  & t & ::= & x & \mbox{ (variable) } \\
D & ::= & f(\ol{x_{n}}) = e & \mbox{ (rule) } 
  & & | & c(\ol{t_{n}}) & \mbox{ (constructor call) } \\
e & ::= & t & \mbox{ (term) } 
  & & | & f(\ol{t_{n}}) & \mbox{ (function call) } \\
  & | & \caseof{x} \{\ol{p_{m}\to e_{m}}\} & \mbox{ (rigid case) } 
  & p & ::= & c(\ol{x_{n}}) & \mbox{ (flat pattern) } \\
  & | & \fcaseof{x} \{\ol{p_{m}\to e_{m}}\} & \mbox{ (flexible case) } \\
\end{array}
\]
\rule{\linewidth}{.01in}
\caption{Syntax of Flat Programs}
\label{flat}
\end{figure}
We consider the following domains:
\[
\begin{array}{l@{~~}l@{~~}l@{~~~}l@{~~~}l@{~~}l@{~~}l@{~~~}l}
x,y,z & \in & \cX & \mbox{\sf (variables)} &
a,b,c & \in & \cC & \mbox{\sf (constructor symbols)} \\
f,g,h & \in & \cF & \mbox{\sf (defined functions)} &
e_{1},e_{2},\ldots & \in  & \cE & \mbox{\sf (expressions)} \\
t_{1},t_{2},\ldots & \in  & \cT & \mbox{\sf (terms)} &
v_{1},v_{2},\ldots & \in  & \cV & \mbox{\sf (values)}
\end{array}
\]
The only difference between \emph{terms} and \emph{expressions} is
that the latter may contain case expressions.  
\emph{Values} are terms in head normal form, i.e., variables or
constructor-rooted terms.  A program $\cR$ consists of a sequence of
function definitions; each function is defined by a single rule whose
left-hand side contains only different variables as parameters.  The
right-hand side is an expression $e$ composed by variables,
constructors, function calls, and case expressions for
pattern-matching.  The general form of a case expression is:\footnote{
  We write $\foo{(f)case}$ for either $\foo{fcase}$ or $\foo{case}$.}
\[
\foo{(f)case}\;x\;\foo{of}\;\{c_1(\ol{x_{n_1}}) \to
e_1;\ldots;c_m(\ol{x_{n_m}}) \to e_m\}
\]
where $x$ is a variable, $c_1,\ldots,c_m$ are different constructors
of the type of $x$, and $e_1,\ldots, e_m$ are expressions (possibly
containing nested $\foo{(f)case}$'s).  The variables $\ol{x_{n_i}}$
are local variables which occur only in the corresponding
subexpression $e_i$.  The difference between $\foo{case}$ and
$\foo{fcase}$ only shows up when the argument, $x$, is a free
variable (within a particular computation): $\foo{case}$
suspends---which corresponds to \emph{residuation}, i.e., pure
functional reduction---whereas $\foo{fcase}$ nondeterministically
binds this variable to a pattern in a branch of the case
expression---which corresponds to either narrowing \cite{AEH00} and
driving \cite{Tur86b}. Note that our functional logic language mainly
differs from typical (lazy) functional languages in the presence of
flexible case expressions.

\begin{example} \label{app-ex}
  Consider again the rules defining functions ``\pr{$+$}''
  (Example~\ref{ex1}) and ``\pr{$\sleq$}'' (Example~\ref{ex-leq}).
  These functions can be defined in the flat representation as
  follows:\footnote{ Although we consider in this work a first-order
    representation---the flat language---we use a curried notation in
    concrete examples (as in Curry). }
\startprog
x $+$ y = fcase x of $\{$ Zero   $\to$ y;
                     $\:$Succ n $\to$ Succ (n $+$ y) $\}$\\[-1ex]

x $\sleq$ y = fcase x of $\{$ Zero   $\to$ True;
                     $\:$Succ n $\to$ fcase y of $\{$ Zero   $\to$ False;
                                            $\:$Succ m $\to$ n $\sleq$ m $\}$ $\}$
\stopprog
\end{example}
An automatic transformation from source (inductively sequential)
programs to flat programs has been introduced by Hanus and Prehofer
\citeyear{HP99}.  Translated programs always fulfill the following
restrictions: case expressions in the right-hand sides of program
rules appear always in the outermost positions (i.e., there is no case
expression inside a function or constructor call) and all case
arguments are variables, thus the syntax of Fig.~\ref{flat} is general
enough for our purposes.  We shall assume these restrictions on flat
programs in the following.

The operational semantics of flat programs is shown in
Fig.~\ref{standard-lnt}. It is based on the LNT---for Lazy Narrowing
with definitional Trees---calculus of Hanus and Prehofer
\citeyear{HP99}. The one-step transition relation $\To_\sigma$ is
labeled with the substitution $\sigma$ computed in the step. Let us
briefly describe the LNT rules:

The {\sf select} rule selects the appropriate branch of a case
expression and continues with the evaluation of this branch.  This
rule implements \emph{pattern matching}.
  
\begin{figure}[t]
\centering\small
\rule{\linewidth}{.01in}
\[
\mbox{}\hspace{-5ex}\begin{array}{l@{~~~~~}rcl}
\mbox{\sf (select)} & \foo{(f)case}\;c(\ol{t_n})\;of\;\{\ol{p_m \to
    e_m}\} & \To_{\id} & \sigma(e_i) \\
& \multicolumn{3}{r}{\mbox{if $p_i = c(\ol{x_n})$, $c\in \cC$,
and $\sigma = \{\ol{x_n \mapsto t_n}\}$}}\\[1ex]
\mbox{\sf (guess)} & \foo{fcase}\;x\;of\;\{\ol{p_m \to e_m}\} 
& \To_{\sigma} & \sigma(e_i) \\
& \multicolumn{3}{r}{\mbox{if $\sigma = \{x \mapsto p_i\}$ 
and $i \in \{1,\ldots,m\}$}}\\[1ex]
\mbox{\sf (case eval)} & \foo{(f)case}\;e\;of\;\{\ol{p_m \to e_m}\} 
& \To_\sigma & \sigma((f)case\;e'\;of\;\{\ol{p_m \to e_m}\}) \\
& \multicolumn{3}{r}{\mbox{if $e$ is not in head normal form and $e \To_{\sigma} e'$}}\\[1ex]
\mbox{\sf (fun)} & f(\ol{t_n}) & \To_{\id} & \sigma(e) \\
& \multicolumn{3}{r}{\mbox{if $f(\ol{x_n}) = e \in \cR$
  and $\sigma = \{\ol{x_n \mapsto t_n}\}$}}
\end{array}
\]
\rule{\linewidth}{.01in}
\caption{Standard Operational Semantics (LNT calculus)} 
\label{standard-lnt}
\end{figure}

The {\sf guess} rule applies when the argument of a \emph{flexible}
case expression is a variable. Then, this rule non-deterministically
binds this variable to a pattern in a branch of the case expression.
The step is labeled with the computed binding.  Observe that there is
no rule to evaluate a rigid case expression with a variable argument.
This situation produces a \emph{suspension} of the evaluation.

The {\sf case eval} rule can be applied when the argument of the case
construct is not in head normal form (i.e., it is either a function
call or another case construct). Then, it tries to evaluate this
expression recursively.

Finally, the {\sf fun} rule performs the unfolding of a function call.
As in proof procedures for logic programming, we assume that we take a
program rule with fresh variables in each such evaluation step.
  
Note that there is no rule to evaluate terms in head normal form; in
this case, the computation stops \emph{successfully}.  An LNT
\emph{derivation} is denoted by $e_0 \To^\ast_\sigma e_n$, which is a
shorthand for the sequence $e_0 \To_{\sigma_1} \ldots \To_{\sigma_n}
e_n$ with $\sigma = \sigma_n \circ \cdots \circ \sigma_1$ (if $n=0$
then $\sigma = id$).  An LNT derivation $e \To^\ast_\sigma e'$ is
\emph{successful} when $e'$ is in head normal form. Then, we say that
$e$ \emph{evaluates} to $e'$ with \emph{computed answer} $\sigma$.

\begin{example} \label{app-der}
  Consider the function ``\pr{$\sleq$}'' of Example~\ref{app-ex}.
  Given the initial call ``\pr{(Succ x) $\sleq$ y}'', the LNT calculus
  computes, among others, the following successful derivation:\\[1ex]
$
\begin{array}{l}
\tt (Succ\;x) \;\sleq\; y \\
\begin{array}{llr}
 \To_{id} & \tt \!\!fcase\;(Succ\;x)\;of\; & ({\sf fun})\\
& \begin{array}{lll}
  \tt \{ Z & \!\!\!\!\to\!\!\!\! & \tt True; \\
  \tt (Succ\;n) & \!\!\!\!\to\!\!\!\! & \tt fcase\;y\;of\;\{ Z \to False;\;(Succ\;m) \to n \sleq m \}\}
\end{array}\\
 \To_{id} & \tt \!\!fcase\;y\;of\;\{ Z \to \tt False;\; (Succ\;m) \to \tt x \sleq m \}& ({\sf select})\\
 \To_{\tt \{y \mapsto Z\}} & \tt \!\!False & ({\sf guess})\\
\end{array}
\end{array}
$\\[1ex]
Therefore, \pr{(Succ x) $\sleq$ y} evaluates to \pr{False} with
computed answer \pr{$\{$y $\mapsto$ Z$\}$}.
\end{example}

\section{Forward Slicing}
\label{pe}

In this section, we formalize our notion of \emph{forward slice} in
the context of functional logic programs. As mentioned before, in our
setting any function call may play the role of \emph{slicing
  criterion}. Essentially, given a program $\cR$ and a (partially
instantiated) call $t$|the slicing criterion|an associated forward
slice is a \emph{fragment} of $\cR$ which contains all the statements
which are necessary for executing the call $t$, i.e., which are
\emph{needed} to evaluate the slicing criterion. This relation between
neededness|in the sense of Huet and L\'evy \citeyear{HL92}|and slicing
is not new; indeed, there exist several approaches to slicing of
functional programs which rely on the computation of neededness
information \cite{Bis97,FT98}. Clearly, $t$ must compute the same
value in $\cR$ and in the computed slice. In particular, the original
program is always a correct slice w.r.t.\ any slicing criterion. Our
aim is thus to find smaller slices.\footnote{Weiser proved that
  computing the \emph{minimal} slice is generally undecidable
  \cite{Wei84}.} Furthermore, we do not distinguish between dynamic
and static slicing, since it only depends on the degree of
instantiation of the slicing criterion; in order words, we consider a
sort of quasi static slicing \cite{Ven91}.

As mentioned before, we do not consider the construction of amorphous
slices;
otherwise, partial evaluation could straightforwardly be seen as a
slicing technique. Here, we only allow the deletion of some elements
of the original program:
\begin{description}
\item[\sf Term deletion:] This is the simplest kind of deletion. It
  consists of the removal of subterms which are not needed to perform
  computations with the slicing criterion.
  
\item[\sf Branch deletion:] By using the partially known data in the
  slicing criterion, some case branches become useless and can be
  deleted.
  
\item[\sf Function deletion:] Finally, those functions which are not
  necessary to evaluate the slicing criterion can be completely
  deleted from the slice.
\end{description}
Analogously to Schoening and Ducass\'e \citeyear{SD96}, our notion of
\emph{program slice} is formalized in terms of an abstraction
relation. In the following, we consider that program signatures are
implicitly augmented with the 0-ary constructor $\top$, a special
symbol which is used to denote that some code fragment is missing.

\begin{definition}[term abstraction]
  A term $t'$ is an abstraction of term $t$, in symbols $t' \succeq
  t$, iff $t' = \top$ or $t' = t$.
\end{definition}

\begin{definition}[expression abstraction]
  An expression $e'$ is an abstraction of an expression $e$, in
  symbols $e' \succeq e$, iff one of the following conditions holds:
  \begin{itemize}
  \item $e' = \top$ (i.e., a case structure is completely deleted);
  \item $e' = e$;
  \item $e' = \foo{(f)case}\;x\;\foo{of}\;\{\ol{p'_n \to
      e'_n}\}$, $e = \foo{(f)case}\;x\;\foo{of}\;\{\ol{p_n \to
      e_n}\}$, and $e'_i \succeq e_i$ for all $i = 1,\ldots,n$.
  \end{itemize}
\end{definition}

\begin{definition}[program slice] \label{prog-abs}
  A program $\cR' = (D'_1,\ldots,D'_m)$ is a slice of a program $\cR =
  (D_1,\ldots,D_m)$, in symbols $\cR' \succeq \cR$, iff for all $i =
  1,\ldots,m$, $D'_i = (f(\ol{x_n}) = e')$, $D_i = (f(\ol{x_n}) = e)$,
  and $e' \succeq e$.
\end{definition}
Roughly speaking, a program $\cR'$ is a slice of program $\cR$ if it
can be obtained by replacing some subterms, case branches, and
right-hand sides of function definitions by $\top$. Trivially, program
slices are steadily executable (and fulfill the syntax of
Fig.~\ref{flat}) by just considering $\top$ as an arbitrary constant
of the program's signature. The interest in producing
\emph{executable} slices comes from the fact that it facilitates
program reuse and, more importantly, it allows us to apply a number of
existing techniques to the computed slice (e.g., debugging, program
analysis, verification, program transformation).

So far, we have only considered the \emph{shape} of a slice. Now, we
consider the \emph{semantics} of the slicing process:

\begin{definition}[correct slice] \label{slice-def}
  Let $\cR$ be a program and $t$ a term. We say that $\cR'$ is a
  correct slice of $\cR$ w.r.t.\ $t$ iff
  \begin{itemize}
  \item $\cR'$ is a program slice of $\cR$ (i.e., $\cR' \succeq \cR$),
    and
  \item $t \To^\ast_{\sigma_1} t_1$ in $\cR$ iff $t
    \To^\ast_{\sigma_2} t_2$ in $\cR'$, where $t_1,t_2$ are values
    (different from $\top$), $t_2 \succeq t_1$, and $\sigma_1 =
    \sigma_2$ (modulo variable renaming).
  \end{itemize}
\end{definition}
Observe that evaluations in the slice may produce values with some
occurrences of $\top$ at inner positions, which is safe in our context
since only the outermost symbol is \emph{observable} in the LNT
semantics. On the other hand, no abstraction is needed for
substitutions, since the computed bindings can only map variables to
patterns of the form $c(\ol{x_n})$ with no occurrences of $\top$ (see
rule {\sf guess} in Fig.~\ref{standard-lnt}).

\begin{figure}[t]
\rule{\linewidth}{.01in}
\begin{flushleft}
\vspace{-3ex}
\startprog
  main op xs = fcase op of $\{$ Len $\to$ fst (lenmax xs);
                           $\:$Max $\to$ snd (lenmax xs) $\}$

lenmax xs = (len xs, max xs)

len xs = fcase xs of $\{$ [] $\to$ Zero;
                   (x:xs) $\to$ Succ (len xs) $\}$

max xs = fcase xs of $\{$ (y:ys) $\to$ fcase ys of 
                                   $\{$ [] $\to$ y;
                                 (z:zs) $\to$ if (y $\sleq$ z) then max (z:zs)
                                                       $\!$else max (y:zs) $\}$ $\}$

x $\sleq$ y = fcase x of $\{$ Zero   $\to$ True;
                     $\:$Succ n $\to$ fcase y of $\{$ Zero   $\to$ False;
                                            $\:$Succ m $\to$ n $\sleq$ m $\}$ $\}$

fst x = fcase x of $\{$ (a,b) $\to$ a $\}$

snd x = fcase x of $\{$ (a,b) $\to$ b $\}$
\stopprog
\end{flushleft}
\vspace{-3ex}
\rule{\linewidth}{.01in}
\caption{Example \pr{lenmax}}
\label{ex-lenmax}
\end{figure}

\begin{example} \label{lenapp}
  Consider the program excerpt shown in Fig.~\ref{ex-lenmax} for
  computing the length or the maximum of a list, depending on the
  value of the first parameter of \pr{main}. Standard functions
  ``\pr{len}'', ``\pr{max}'', ``\pr{fst}'', and ``\pr{snd}'' return
  the length of a list, the maximum of a list, the first element of a
  tuple, and the second element of a tuple, respectively.  Given the
  slicing criterion ``\pr{main Len xs}'', the following slice can be
  obtained:
\startprog
main op xs = fcase op of $\{$ Len $\to$ fst (lenmax xs);
                           Max $\to$ $\top$ $\}$

lenmax xs = (len xs, $\top$)

len xs = fcase xs of $\{$ []     $\to$ Zero;
                       (x:xs) $\to$ Succ (len xs) $\}$

max xs = $\top$

x $\sleq$ y = $\top$

fst x = fcase x of $\{$ (a,b) $\to$ a $\}$

snd x = $\top$
\stopprog
Here, we have performed three different kinds of code deletion:
\begin{description}
\item[\sf Term deletion:] The evaluation of the call to function
  ``\pr{max}'' in the right-hand side of ``\pr{lenmax}'' is not
  needed|since function ``\pr{fst}'' only demands the evaluation of
  the first component of the tuple|and, thus, it has been replaced by
  $\top$.
  
\item[\sf Branch deletion:] In the definition of function
  ``\pr{main}'', the second branch of the case expression is not
  needed to execute the slicing criterion; therefore, it has also been
  replaced by $\top$.
  
\item[\sf Function deletion:] Since functions ``\pr{max}'',
  ``\pr{$\sleq$}'', and ``\pr{snd}'' are no longer necessary to
  evaluate the slicing criterion, their definitions have been replaced
  by $\top$.
\end{description}
Note that this slice could not be constructed by using a simple graph
of functional dependencies (e.g., functions ``\pr{snd}'',
``\pr{lenmax}'', and ``\pr{$\sleq$}'' depend on function ``\pr{main}''
but they do not appear in the computed slice).
\end{example}
In order to simplify the representation of program slices, in the
following we adopt the following conventions:
\begin{itemize}
\item case branches of the form $p \to \top$ are deleted and
\item function definitions of the form $f(\ol{x_n}) = \top$ are
  removed from the slice.
\end{itemize}
Therefore, the slice of Example~\ref{lenapp} is simply written as
follows:
\startprog
main op xs = fcase op of $\{$ Len $\to$ fst (lenmax xs) $\}$

lenmax xs = (len xs, $\top$)

len xs = fcase xs of $\{$ []     $\to$ Zero;
                       (x:xs) $\to$ Succ (len xs) $\}$

fst x = fcase x of $\{$ (a,b) $\to$ a $\}$
\stopprog

\section{Monovariant/Monogenetic Partial Evaluation} \label{pe1}

As discussed in the introduction, our developments rely on the fact
that forward slicing can be regarded as a form of
monovariant/monogenetic partial evaluation. This requirement is
necessary in order to ensure that there is a one-to-one relation
between the functions of the original and residual programs, which is
crucial to produce a fragment of the original program rather than a
specialized version. 

In this section, we first recall the basic narrowing-driven partial
evaluation (NPE) scheme \cite{AV02} and, then, modify it in order to
obtain a monovariant and monogenetic partial evaluator.

Essentially, NPE proceeds by iteratively unfolding a set of function
calls, testing the \emph{closedness} of the unfolded expressions, and
adding to the current set those calls (in the derived expressions)
which are not closed. This process is repeated until all the unfolded
expressions are closed, which guarantees the correctness of the
transformation process \cite{AFV98}, i.e., that the resulting set of
expressions covers all the possible computations for the initial call.
This iterative style of performing partial evaluation was first
described by Gallagher \citeyear{Gal93} for the partial evaluation of
logic programs.

The computation of a closed set of expressions can be regarded as the
construction of a graph containing the program points which are
reachable from the initial call. Intuitively, an expression is
\emph{closed} whenever its maximal operation-rooted subterms (function
calls) are instances of the already partially evaluated terms.
Formally, the closedness condition is defined as follows:

\begin{definition}[closedness] \label{closedness}
Let $E$ be a finite set of expressions. We say that an expression
$e$ is closed w.r.t.\ $E$ (or $E$-closed) iff one of the following 
conditions hold:
\begin{itemize}
\item $e$ is a variable;
\item $e=c(e_{1},\ldots,e_{n})$ is a constructor call
and $e_{1},\ldots,e_{n}$ are recursively $E$-closed; 
\item $e=(f)case\;e'\;of\;\{\ol{p_{m} \to e_{m}}\}$ is a case
  expression and $e',e_{1},\ldots,e_{m}$ are recursively $E$-closed;
\item $e$ is operation-rooted, there is an expression $e' \in E$, a
  matching substitution $\sigma$ with $e = \sigma(e')$, and, for all
  $x \mapsto e'' \in \sigma$, $e''$ is recursively $E$-closed.
\end{itemize}
\end{definition}

\begin{figure}[t]
\centering
\rule{\linewidth}{.01in}
\begin{minipage}{8cm}
\mbox{}\\
{\bf Input:} a program $\cR$ and a term $t$\\
{\bf Output:} a residual program $\cR'$\\
{\bf Initialization:} $i:= 0;~ E_0 := \{t\}$\\
{\bf Repeat} \mbox{}\\
\mbox{}\hspace{3ex} $E' := \foo{unfold}(E_i,\cR)$;\\
\mbox{}\hspace{3ex} $E_{i+1}:= \foo{abstract}(E_i,E')$;\\
\mbox{}\hspace{3ex} $i := i+1$;\\
{\bf Until} $E_i = E_{i-1}$ (modulo renaming)\\
{\bf Return:} \\
\mbox{}\hspace{3ex} $\cR' := \foo{build\_residual\_program}(E_i,\cR)$\\
\end{minipage}
\rule{\linewidth}{.01in}
\caption{Narrowing-Driven Partial Evaluation Procedure}
\label{pe-alg}
\end{figure}

\noindent
The basic partial evaluation procedure is shown in Fig.~\ref{pe-alg}.
Let us explain the operators in this procedure:
\begin{itemize}
\item The operator $\foo{unfold}$ takes a program and a set of
  expressions $E_i = \{e_1,\ldots,e_n\}$, computes a \emph{finite} set
  of (possibly incomplete) finite derivations $e_j \To^\ast e'_j$, $j
  = 1,\ldots,n$, and returns the set of derived expressions $E' =
  \{e'_1,\ldots,e'_n\}$. Here, partial computations are performed with
  the LNT calculus of Fig.~\ref{standard-lnt} slightly extended to
  avoid the backpropagation of bindings: the RLNT (for Residualizing
  LNT) calculus of Albert et al.\ \citeyear{AHV03}. The main
  difference between the LNT and the RLNT calculi is that the
  non-deterministic rule \textsf{guess} of the LNT calculus is
  replaced by a deterministic rule that leaves the case structure
  untouched and proceeds with the evaluation of the branches.
  
\item Function $\foo{abstract}$ is then used to properly add the new
  expressions to the current set of (to be) partially evaluated
  expressions. For instance, a trivial abstraction operator could be
  defined as follows:
  \[
  \foo{abstract}(E_i,E') = E_i \cup \{ e \in E' \mid \mbox{ there is
    no } e'\in E_i \mbox{ such that } e = e' \}
  \]
  Here, only the new expressions that are not equal (modulo variable
  renaming) to some expression in the current set $E_i$ are added.
  This abstraction operator, however, does not guarantee the
  termination of the process since an infinite number of different
  expressions can be derived. In general, a termination test is also
  applied, e.g., Alpuente et al.\ \citeyear{AFV98} consider a variant
  of the Kruskal tree condition called ``homeomorphic embedding''
  \cite{Leu02}: if an expression embeds another expression in the
  current set, some form of generalization|usually the \emph{most
    specific generalization} operator|is applied and the generalized
  term is added to the current set.
  
\item The main loop of the algorithm can be seen as a
  \emph{pre-processing} stage whose aim is to find a closed set of
  expressions.  Note that no residual rules are actually constructed
  during this phase.  Only when a closed set of expressions is
  eventually found, residual rules are built as follows:
  \[
  \foo{build\_residual\_program}(E_i,\cR) = \{ e = e' \mid e \in E_i
    \mbox{ and } e \To^\ast e' \mbox{ in } \cR \}
  \]
  In general, this operator also applies a renaming of expressions and
  some post-unfolding transformations which are not relevant for this
  work; we refer the interested reader to \cite{AV02}.
\end{itemize}
In principle, the NPE scheme has been designed to achieve both
polyvariant and polygenetic specializations. In this work, however, we
are interested in the definition of a less powerful monovariant and
monogenetic scheme.  For this purpose, we should impose several
restrictions to the procedure of Fig.~\ref{pe-alg}:
\begin{enumerate}
\item Firstly, the current set $E_i$ should only contain
  operation-rooted terms without nested function calls (i.e., of the
  form $f(\ol{t_{n}})$, where $f$ is a defined function symbol and
  $t_1,\ldots,t_n$ are constructor terms). This is necessary to ensure
  that partial evaluation is monogenetic and, thus, we do not produce
  residual functions that mix several functions of the original
  program.
 
\item Secondly, the unfolding operator should perform only a one-step
  evaluation of each call rather than a computation of an arbitrary
  length. This condition is required to guarantee that no reachable
  function is hidden by the unfolding process. For instance, if we
  would allow a computation of the form $\pr{f x $\To$ g x $\To$ h
    x}$, the unfolding operator would only return $\pr{h x}$, while
  $\pr{g x}$ should also be part of the slice.
 
\item Finally, the abstraction operator should ensure that the current
  set of terms contains at most one term for each function symbol. In
  this way, we enforce the monovariant nature of the partial
  evaluation process, i.e., that only one residual definition is
  produced (at most) for each original function.
\end{enumerate}
Unfortunately, such a monovariant/monogenetic partial evaluator would
propagate information poorly. In order to overcome this drawback, in
the next section we introduce a carefully designed operational
mechanism which avoids the loss of information (i.e., program
dependences) as much as possible.
  
\section{Computing Program Dependences} \label{depen}

In this section, we introduce the kernel of a monovariant and
monogenetic partial evaluator that can be used to compute program
dependences. In principle, such a partial evaluator could proceed as
follows:
\begin{itemize}
\item terms containing nested function symbols are \emph{flattened};
\item terms in the current set of (to be) partially evaluated terms
  which are rooted by the same function symbol are generalized with
  some appropriate generalization operator (e.g., the \emph{most
    specific generalization} operator).
\end{itemize}
However, flattening terms with nested function symbols would imply a
serious loss of precision. For instance, a term of the form ``\pr{fst
  (lenmax xs)}'' would be replaced by the terms ``\pr{fst y}'' and
``\pr{lenmax xs}'', where \pr{y} is a fresh variable, thus missing the
fact that \pr{fst} is called with the result of ``\pr{lenmax xs}''.

In order to avoid this loss of precision, we drop the first
restriction above, i.e., we consider arbitrary operation-rooted terms
during partial evaluation. However, we should still ensure that only a
one-step of unfolding is applied to each term in order to guarantee
that no reachable function is hidden by the unfolding process. In our
flat language, function calls are evaluated \emph{lazily}: a term
containing nested function calls is evaluated by first unfolding the
outermost function; inner function calls are only evaluated \emph{on
  demand}, i.e., when they appear as the argument of some case
expression.
For instance, ``\pr{fst (lenmax xs)}'' is unfolded to 
\[
\pr{fcase (lenmax xs) of $\{$ (a,b) $\to$ a $\}$}
\]
Then, the evaluation of function ``\pr{fst}'' cannot continue until
the inner call to ``\pr{lenmax}'' is reduced to a value.
Unfortunately, this interleaved evaluation is problematic in our
context since it would give rise to a polygenetic partial evaluation
(i.e., a residual function comprising the evaluation of both \pr{fst}
and \pr{lenmax}) .  In contrast, we should perform a \emph{complete
  one-step unfolding} of each function call separately, i.e., a
function unfolding followed by the reduction of all the case
structures in the unfolded expression.  

For this purpose, we extend the partial evaluation mechanism in order
to work on \emph{states} rather than on expressions.

\begin{definition}[state]
  A state is a pair of the form $\tuple{e,S}$, where $e$ is an
  expression (to be evaluated) and $S$ is a \emph{stack} (a list)
  which represents the current ``evaluation
  context''.\footnote{Similar operational semantics with a stack can
    be found in \cite{AHHOV05,Ses97}.}  The empty stack is denoted by
  $\nil$.
\end{definition}
For example, the previous expression ``\pr{fst (lenmax xs)}'' could
now be flattened as follows (see Example~\ref{ex-flat}): $\tt
\tuple{\pr{lenmax xs},~[(\pr{fst x},x)]}$, which means that
$\pr{lenmax xs}$ is ready to perform a complete one-step unfolding;
when this evaluation is performed, the initial term can be
reconstructed thanks to the information in the stack, $\tt (\pr{fst
  x},x)$, which means that the initial term has the form $\pr{fst x}$,
where \pr{x} is the result of evaluating the first component of the
state (i.e., the result of evaluating \pr{lenmax xs}). Thanks to the
use of states, we do not miss the fact that \pr{fst} is called with
the result of ``\pr{lenmax xs}''.

Figure~\ref{smallstep} shows an extended operational semantics which
is appropriate to deal with states. Let us briefly explain the rules
of this operational semantics.

Rules {\sf select} and {\sf guess} proceed in a similar way as their
counterpart in the standard semantics of Fig.~\ref{standard-lnt}.

Rule {\sf flatten} is used to avoid the unfolding of those
(operation-rooted) terms whose unfolding would demand the evaluation
of some inner call. This is necessary to ensure that partial
evaluation is monogenetic. In this case, we delay the function
unfolding and continue with the evaluation of the demanded inner call.
Auxiliary function \emph{flat} is used to flatten these states. Here,
we use subscripts in the arrows to indicate the application of some
concrete rule(s).  Function \emph{flat} proceeds as follows:
\begin{itemize}
\item[] When the expression in the input state can be reduced by using
  rules {\sf select} and {\sf guess} to a case expression with a
  function call in the argument position (which is thus
  \emph{demanded}), function \emph{flat} returns a new state whose
  first component is the demanded call, $g(\ol{t'_{m}})$, and whose
  stack is augmented by adding a new pair
  $(f(\ol{t_{n}})[g(\ol{t'_{m}})/x], x)$.  Here,
  $f(\ol{t_{n}})[g(\ol{t'_{m}})/x]$ denotes the term obtained from
  $f(\ol{t_{n}})$ by replacing the selected occurrence of the inner
  call, $g(\ol{t'_{m}})$, with a fresh variable $x$.  This pair
  contains all the necessary information to reconstruct the original
  expression once the inner call is evaluated to a value (in rule {\sf
    replace}).
\end{itemize}
\begin{example} \label{ex-flat}
  Consider again the program of Example~\ref{lenapp}. In order to
  flatten the following expression: $\tt \tuple{\pr{fst (lenmax
      xs)},~\pr{[]}}$, we proceed as follows. First, we perform a
  function unfolding so that we get:
  \[
  \tt \tuple{\pr{fcase (lenmax xs) of } \{ \pr{(a,b)} \to \pr{a} \},~ \nil}
  \]
  Now, we try to evaluate this state by means of rules \textsf{select}
  and \textsf{guess}. Since no reduction is possible and the case
  structure has a function call in the argument position, function
  $\foo{flat}$ returns the state
  \[
  \tt \tuple{\pr{lenmax xs},~[(\pr{fst x},\pr{x})]}
  \]
  where \pr{x} is a fresh variable. Observe that this state cannot be
  further flattened since a function unfolding returns the state
  \[
  \tt \tuple{(\pr{len xs},~\top), ~\nil }
  \]
  which cannot be reduced by rules \textsf{select} and \textsf{guess}
  and which contains no function call in the argument position of a
  case expression. Therefore, in this case, function $\foo{flat}$
  returns $\bot$ and no step with rule \textsf{flatten} can be done.
\end{example}
Rule {\sf fun} performs a simple function unfolding when rule {\sf
  flatten} does not apply, i.e., when function \emph{flat} returns
$\bot$.

Finally, rule {\sf replace} allows us to retake the evaluation of some
delayed function call once the demanded inner call is reduced to a
value.

\begin{figure}[t]
\rule{\linewidth}{.01in}
\[
\mbox{}\hspace{-5ex}\begin{array}{l@{~~~~~}lr@{~~~\To~~~}lr}
\mbox{\sf (select)} & \l\: \foo{(f)case}\;c(\ol{t_n})\;of\;
        \{\ol{p_{k} \to e_{k}}\}, & S\:\r & \l\:\rho(e_i),  & S\:\r \\[.5ex]
 & \multicolumn{4}{l}{\mbox{if } p_i = c(\ol{x_n}),\;c\in\cC, \mbox{ and } 
                      \rho = \{\ol{x_n \mapsto t_n}\} } \\[1ex]
\mbox{\sf (guess)} & \l\: \foo{(f)case}\;x\;of\;\{\ol{p_{k} \to e_{k}}\}, 
      & S\:\r & \l\:\rho(e_i), & S\:\r  \\[.5ex]
  & \multicolumn{4}{l}{\mbox{if } 
     \rho = \{x \mapsto p_{i}\} \mbox{ and } i \in\{1,\ldots,k\}} \\[1ex]
\mbox{\sf (flatten)} & \l\: f(\ol{t_{n}}), & S\:\r &
  \l\: g(\ol{t'_{m}}), & S^{f}\:\r  \\[.5ex]
  & \multicolumn{4}{l}{\mbox{if } \foo{flat}(f(\ol{t_{n}}),S)= 
  \tuple{g(\ol{t'_{m}}),S^{f}}} \\[1ex]
\mbox{\sf (fun)} & \l\:f(\ol{t_{n}}), & S\:\r &  \l\:\rho(e), & S\:\r \\[.5ex]
& \multicolumn{4}{l}{\mbox{if } \foo{flat}(f(\ol{t_{n}}),S)=\bot,\:
 f(\ol{x_{n}}) = e \in \cR,\: \mbox{ and } 
\rho=\{\ol{x_{n} \to t_{n}}\} } \\[1ex]
\mbox{\sf (replace)} & \l\: v, & (f(\ol{t_{n}}),x):S\:\r  
& \l\:\rho(f(\ol{t_{n}})), & S\:\r  \\[.5ex]
& \multicolumn{4}{l}{\mbox{if }  v \mbox{ is a value and } 
  \rho=\{x \mapsto v\}}  
\end{array}
\]
\[
\mbox{}\hspace{-5ex}\begin{array}{lll}
\mbox{\sf where}\;\;\foo{flat}(\tuple{f(\ol{t_{n}}),S}) & \;=\; &
\mbox{\sf if } \tuple{\rho(e),\nil} \:\Longrightarrow^\ast_{\sf select/guess}\:
\tuple{\foo{(f)case}\;g(\ol{t'_{m}})\;of\;\{\;\ldots\;\},\nil} \\
& & \mbox{}\hspace{3em}\mbox{\sf then }\; 
\tuple{g(\ol{t'_{m}}),(f(\ol{t_{n}})[g(\ol{t'_{m}})/x],
 x):S}\\
& & \mbox{}\hspace{3em}\mbox{\sf else }\;\: \bot \\ 
& & \mbox{\sf with } f(\ol{x_{n}}) = e \in \cR, \;\mbox{\sf and } \rho = \{\ol{x_{n} \to t_{n}}\}
\end{array}
\]
\rule{\linewidth}{.01in}
\caption{Extended Operational Semantics}
\label{smallstep}
\end{figure}

The extended operational semantics behaves almost identically to the
standard semantics of Fig.~\ref{standard-lnt}.  There are, though, the
following main differences:
\begin{itemize}
\item Now, the one-step relation $\To$ is not labeled with the
  computed bindings since we are not interested in computing answers
  but only in obtaining the functions which are reachable from the
  initial call.
  
\item In the standard semantics, rigid case expressions with a free
  variable in the argument position \emph{suspend}.  In our case, rule
  {\sf guess} proceeds with their evaluation as if they were flexible.
  This is motivated by the fact that we may have \emph{incomplete}
  information; hence, in order to be on the safe side|and do not miss
  any reachable function|we should explore all the alternatives of
  rigid case expressions.
  
\item The order of evaluation is slightly changed. In our extended
  semantics, we delay those function unfoldings which cannot be
  followed by the reduction of all the case expressions in the
  corresponding right-hand side.
\end{itemize}
In spite of these differences, both calculi trivially produce the same
results for input expressions involving no suspension.  Roughly
speaking, the extended semantics is in between the standard
operational semantics and its residualizing version used to perform
partial computations in the NPE framework \cite{AHV03}.

\begin{example} \label{ex-comps}
  Consider again the program of Example~\ref{lenapp}. Given the
  initial term $\pr{fst (lenmax xs)}$, we have (among others) the
  following (incomplete) computation with the standard semantics of
  Fig.~\ref{standard-lnt}:
  \[ \small
  \begin{array}{llll}
    \pr{fst (lenmax xs)} 
     & \tt \To_{id} & \pr{fcase (lenmax xs) of } \{ \pr{(a,b)} \to \pr{a} \} & \mbox{(\textsf{fun})}\\
     & \tt \To_{id} & \pr{fcase (len xs, max xs) of } \{ \pr{(a,b)} \to \pr{a} \} & \mbox{(\textsf{fun})}\\
     & \tt \To_{id} & \pr{len xs} & \mbox{(\textsf{select})}
  \end{array}
  \]
  On the other hand, the extended operational semantics of
  Fig.~\ref{smallstep} performs the following equivalent derivation:
  \[ \small \hspace{-8ex}
  \begin{array}{llll}
    \tuple{\pr{fst (lenmax xs)},~\nil}
     & \tt \To & \tuple{\pr{lenmax xs},~[(\pr{fst x},\pr{x})]} & \mbox{(\textsf{flatten})}\\
     & \tt \To & \tuple{\pr{(len xs, max xs)},~[(\pr{fst x},\pr{x})]} & \mbox{(\textsf{fun})}\\
     & \tt \To & \tuple{\pr{fst (len xs, max xs)},~\nil} & \mbox{(\textsf{replace})}\\
     & \tt \To & \tuple{\pr{fcase (len xs, max xs) of } \{ \pr{(a,b)} \to \pr{a} \},~\nil} & \mbox{(\textsf{fun})}\\
     & \tt \To & \tuple{\pr{len xs},~\nil} & \mbox{(\textsf{select})}\\
  \end{array}
  \]
\end{example}
The relevance of the extended semantics stems from the fact that
computations can now be split into a number of consecutive sequences
of steps of the form:\\[2ex]
$ \mbox{}\hspace{2ex}\Longrightarrow_{\sf flatten}^\ast \underbrace{
  \Longrightarrow_{\sf fun} \Longrightarrow^\ast_{\sf select/guess}
}_{seq\_1} \Longrightarrow^\ast_{\sf replace} \Longrightarrow_{\sf
  flatten}^\ast \underbrace{ \Longrightarrow_{\sf fun}
  \Longrightarrow^\ast_{\sf select/guess} }_{seq\_2}
\Longrightarrow^\ast_{\sf replace} \ldots
$\\[2ex]
where each subsequence, \emph{seq\_i}, represents a complete one-step
unfolding of some function call. From these sequences, a
monogenetic/monovariant partial evaluation scheme can easily be
defined and, thus, the algorithm for computing dependences in our
program slicing technique.  

The algorithm of Fig.~\ref{pe-alg} is now slightly modified in order
to work with states.  The new algorithm (depicted in
Fig.~\ref{pe-alg2}) does not compute a residual program but only the
set of states which are reachable from the initial call. In other
words, it returns the counterpart of the final set of \emph{closed}
terms computed by the algorithm of Fig.~\ref{pe-alg}.  The new
algorithm starts by flattening the initial term in order to ensure
that a complete one-step unfolding can be performed.  We now tackle
the definition of appropriate unfolding and abstraction operators.
First, the \emph{one-step} unfolding operator is defined as follows:

\begin{figure}[t]
\centering
\rule{\linewidth}{.01in}
\begin{minipage}{11cm}
\mbox{}\\
{\bf Input:} a program $\cR$ and an operation-rooted term $t$\\
{\bf Output:} a set of states $\cS$ \\ 
{\bf Initialization:} $i:= 0;~ \cS_0 := \{\tuple{t',S}\}$, 
where $\tuple{t,\nil} \Longrightarrow^\ast_{\sf flatten} \tuple{t',S} 
\not\Longrightarrow_{\sf flatten}$\\
{\bf Repeat} \mbox{}\\
\mbox{}\hspace{3ex} $\cS' := \foo{unfold}(\cS_i,\cR)$;\\
\mbox{}\hspace{3ex} $\cS_{i+1}:= \foo{abstract}(\cS_i,\cS')$;\\
\mbox{}\hspace{3ex} $i := i+1$;\\
{\bf Until} $\cS_i = \cS_{i-1}$ (modulo renaming)\\
{\bf Return:} 
~$\cS := \cS_{i}$\\[-1ex]
\end{minipage}
\rule{\linewidth}{.01in}
\caption{Computation of Reachable Program Points}
\label{pe-alg2}
\vspace{-2ex}
\end{figure}

\begin{definition}[unfold]
  Let $\cS$ be a set of states. The unfolding operator
  $\foo{unfold}$ is defined by
\[
\begin{array}{l}
\foo{unfold}(\cS) = {\displaystyle \bigcup_{s \in \cS}} \;\foo{unf}(s)
\end{array}
\]
where\\[1ex]
$ \mbox{}\hspace{2ex} \foo{unf}(\tuple{t,S}) = \{ \tuple{t',S} \mid
\tuple{t,S} \Longrightarrow_{\sf fun}
\tuple{t'',S}\Longrightarrow^\ast_{\sf select/guess} \tuple{t',S}
\not\Longrightarrow_{\sf select/guess} \}
$
\end{definition}
This unfolding operator always performs a complete one-step unfolding
of each input expression. The associated stack $S$ remains unchanged
since only rules {\sf flatten} and {\sf replace} can modify the
current stack. Function $\foo{unf}$ returns a \emph{set} of derived
states because of the non-determinism of the underlying operational
semantics.

\begin{example}
  Consider again the program of Example~\ref{lenapp}. We illustrate
  function $\foo{unf}$ by means of some simple examples:
  \[
  \foo{unf}(\tuple{\pr{lenmax xs},~[(\pr{fst x},\pr{x})]}) =
  \tuple{\pr{(len xs, max xs)},~[(\pr{fst x},\pr{x})]}
  \]
  \[
  \foo{unf}(\tuple{\pr{fst (len xs, max xs)},~\nil}) = \tuple{\pr{len
      xs},~\nil}
  \]
  according to the partial computation in Example~\ref{ex-comps}.
\end{example}
Before defining our abstraction operator, we need the following
auxiliary notion:

\begin{definition}[flattened state]
  Let $s$ be a state returned by the operator $\foo{unfold}$ with
  $s \Longrightarrow^\ast_{\sf replace/flatten} s'
  \not\Longrightarrow$.  Then $s'$ is called a flattened state.
\end{definition}
Flattened states have a particular form, as stated by the following
result:

\begin{lemma} \label{flattened-states}
Let $s$ be a flattened state. Then $s$ has the form 
$\tuple{v,\nil}$, where $v$ is a value, or $\tuple{f(\ol{t_{n}}),S}$, 
where $f(\ol{t_{n}})$ is an operation-rooted term.
\end{lemma}
In order to add new states to the current set of states, we introduce
the following abstraction operator:

\begin{definition}[abstract]
  Let $\cS$ and $\cS' = \{s_{1},\ldots,s_{n}\}$ be sets of states. Our
  abstraction operator proceeds as follows:
$
\foo{abstract}(\cS,\cS') = 
  \foo{abs}(\foo{abs}(\ldots \foo{abs}(\cS,s'_{1})\ldots, 
  s'_{n-1}),s'_{n})
$,
where:
\[
s_{i} \Longrightarrow^\ast_{\sf replace/flatten} s'_{i} 
\not\Longrightarrow_{\sf replace/flatten} 
~~\mbox{(for all $i = 1,\ldots,n$) }
\]
\end{definition}
Basically, function $\foo{abstract}$ starts by flattening the input
states by applying (zero or one step of) rule {\sf replace}, followed
by (zero or more steps of) rule {\sf flatten}.

\begin{definition}[abs]
  Function $\foo{abs}$ is defined inductively on the structure of
  flattened states (according to Lemma~\ref{flattened-states}):
\begin{description}
\item[$\foo{abs}(\cS,\tuple{x,\nil}) = \cS$] \mbox{}\\[-2ex]

\item[$\foo{abs}(\cS,\tuple{c(\ol{t_{n}}),\nil}) = 
\foo{abstract}(\cS,\cS')$] 
\mbox{}\\
{\sf if $\ol{t'_{m}}$ are the maximal operation-rooted subterms of 
$c(\ol{t_{n}})$ and $\cS' = \{ \ol{\tuple{t'_{m},\nil}}\}$ }\\[-2ex]

\item[$\foo{abs}(\cS,\tuple{f(\ol{t_{n}}),S}) = $] 
\mbox{}\\
$\left\{
\begin{array}{l@{~~~~}l}
\cS \cup \{\tuple{f(\ol{t_{n}}),S}\} & \mbox{\sf if there is no state 
$\tuple{f(\ol{t'_{n}}),S'}$ in $\cS$ } \\
\cS & \mbox{\sf else if $\tuple{f(\ol{t_{n}}),S}$ is $\cS$-closed } \\
\foo{abstract}(\cS^\ast,\cS'') & \mbox{\sf otherwise, where 
$\tuple{f(\ol{t'_{n}}),S'} \in \cS$,} \\ 
& \mbox{\sf $\foo{msg}(\tuple{f(\ol{t'_{n}}),S'},
\tuple{f(\ol{t_{n}}),S}) = (\tuple{f(\ol{t''_{n}}),S'},\cS'')$, } \\
& \mbox{\sf and
$\cS^\ast = (\cS \:\backslash\: \{ \tuple{f(\ol{t'_{n}}),S'}\}) \cup 
\{\tuple{f(\ol{t''_{n}}),S'}\}$ }
\end{array}
\right.
$
\end{description}
\end{definition}
Informally speaking, function $\foo{abs}$ determines the corresponding
action depending on the first component of the new state.  If it is a
variable, we discard the state.  If it is constructor-rooted, we try
to (recursively) add the maximal operation-rooted subterms.  If it is
a function call, then we have three possibilities:
\begin{itemize}
\item If there is no call to the same function in the current set, the
  new state is added to the current set of states.
  
\item If there is a call to the same function in the current set, but
  the new call is \emph{closed} w.r.t.\ this set, it is discarded.
  
\item Otherwise, we generalize the new state and the existing state
  with the same outermost function|which is trivially unique by
  definition of $\foo{abstract}$|and, then, we try to (recursively)
  add the states computed by function $\foo{msg}$.
\end{itemize}
The notion of closedness is easily extended from expressions to
states: a state $\tuple{t,S}$ is \emph{closed} w.r.t.\ a set of states
$\cS$ iff $S[t]$ is $T$-closed (according to Def.~\ref{closedness}),
with $T = \{ S'[t'] \mid \tuple{t',S'} \in \cS \}$.  Here, $S[t]$
denotes the term represented by $\tuple{t,S}$, i.e., inner calls are
moved back to their positions in the outer calls of the stack.  For
instance, given the state 
\[
\tuple{t,S} = \tt \tuple{y, [(len\;x_{2},\; x_{2}),
  (fst\;(x_{1},\;snd\;z), \;x_{1})] }
\]
we have $S[t] = \tt fst\;(len\;y,\;snd\;z)$.

The operator \emph{msg} on states is defined as follows. First, we
recall the standard notion of \emph{msg} on terms: a term $t$ is a
\emph{generalization} of terms $t_{1}$ and $t_{2}$ if both $t_{1}$
and $t_{2}$ are instances of $t$; furthermore, term $t$ is the
\emph{msg} of $t_{1}$ and $t_{2}$ if $t$ is a generalization of
$t_{1}$ and $t_{2}$ and, for any other generalization $t'$ of $t_{1}$
and $t_{2}$, $t$ is an instance of $t'$. Now, the \emph{msg} of two
states is defined by
\[ 
\foo{msg}(\tuple{t_{1},S_{1}}, \tuple{t_{2},S_{2}})
~=~(\tuple{t,S_{1}}, \foo{calls}(\sigma_{1}) \cup
\foo{calls}(\sigma_{2}) \cup \foo{calls}(S_{2}))
\]
where $\foo{msg}(t_{1},t_{2}) = t$, and $\sigma_{1}$ and
$\sigma_{2}$ are the matching substitutions, i.e., $\sigma_{1}(t) =
t_{1}$ and $\sigma_{2}(t) = t_{2}$.  The auxiliary function
$\foo{calls}$ returns a set of states of the form $\tuple{t,\nil}$
for each maximal operation-rooted term $t$ in (the range of) a
substitution or in a stack. 

\begin{example}
  Consider the set of states $\cS = \{ \tuple{\pr{len
      xs},\nil},~\tuple{\pr{fst (a,b)},~\nil}\}$. We illustrate
  function $\foo(abs)$ by means of some simple examples: 
  \[
  \foo{abs}(\cS,\tuple{\pr{max xs},~\nil}) = \cS \cup \{\tuple{\pr{max
      xs},~\nil}\}
  \]
  since there is no state rooted by function \pr{max} in $\cS$,
  \[
  \foo{abs}(\cS,\tuple{\pr{len (y:ys)},~\nil}) = \cS
  \]
  since $\tuple{\pr{len (y:ys)},~\nil}$ is $\cS$-closed (i.e., \pr{len
    (y:ys)} is an instance of \pr{len xs}), and
  \[
  \foo{abs}(\cS,\tuple{\pr{fst z},~\nil}) = \{ \tuple{\pr{len
      xs},\nil},~\tuple{\pr{fst w},~\nil}\}
  \]
  since there is a state $\tuple{\pr{fst (a,b)},\nil}$ rooted by
  function \pr{fst}, the state $\tuple{\pr{fst z},~\nil}$ is not
  $\cS$-closed (since \pr{fst z} is not an instance of \pr{fst (a,b)},
  the most specific generalization of the states $\tuple{\pr{fst
      (a,b)}}$ and $\tuple{\pr{fst z},~\nil}$ returns $\tuple{\pr{fst
      w},~\nil}$, and
  \[
  \foo{abs}(\{ \tuple{\pr{len xs},\nil}\}, \tuple{\pr{fst w},~\nil}) =
  \{ \tuple{\pr{len xs},\nil},~ \tuple{\pr{fst w},~\nil}\}
  \]
  since there is no state in $\{ \tuple{\pr{len xs},\nil}\}$ rooted by
  function \pr{fst}.
\end{example}
Our operator $\foo{abstract}$ can be seen as an instance of the
parametric abstraction operator introduced by Alpuente et al.\ 
\citeyear{AFV98} particularized to consider states (rather than terms)
and monovariant partial evaluation (thus, only one operation-rooted
term is allowed for each defined function symbol). Our abstraction
operator is safe in the following sense:

\begin{lemma} \label{safe-abstract}
  Let $\cS$ be a set of flattened states and $\cS'$ a set of unfolded
  states (as returned by $\foo{unfold}$). Then the states in $\cS \cup
  \cS'$ are closed w.r.t.\ $\foo{abstract}(\cS,\cS')$.
\end{lemma}
This lemma is a crucial result to ensure the correctness of our
approach. In fact, it will allow us to prove that the generated
program is a correct slice according to Definition~\ref{slice-def}.

\begin{example} \label{lenapp2}
  Consider again the program of Example~\ref{lenapp}. Given the
  slicing criterion ``\pr{main Len xs}'', the initial set of states is
  $\tt \cS_{0} = \{\tuple{\pr{main Len xs},\nil} \} $.  Now, we show
  the sequence of iterations performed by the algorithm of
  Fig.~\ref{pe-alg2}:
  \[
  \begin{array}{l}
    \cS'_0 = \{ \tuple{\pr{fst (lenmax xs)},\nil}  \} \\
    \cS_1 = \cS_0 \cup \{ \tuple{\pr{lenmax xs},\;[(\pr{fst x},\pr{x})]} \} \\

    \cS'_1 = \cS'_0 \cup \{ \tuple{(\pr{len xs},\pr{max xs}), \;[(\pr{fst x},\pr{x})] } \} \\
    \cS_2 = \cS_1 \cup \{ \tuple{\pr{fst (len xs, max xs)},\nil} \} \\
    \cS'_2 = \cS'_1 \cup \{ \tuple{\pr{len xs},\nil} \} \\
    \cS_3  = \cS_2 \cup \{ \tuple{\pr{len xs},\nil} \} \\
    \cS'_3 = \cS'_2 \cup \{ \tuple{\pr{Zero},\nil},\;\tuple{\pr{Succ (len xs)},\nil}  \} \\
    \cS_4 = \cS_3
  \end{array}
  \]
  where $\cS'_{i} = \foo{unfold}(\cS_{i},\cR)$ and $\cS_{i+1} =
  \foo{abstract}(\cS_{i},\cS'_{i})$, for $i = 0,\ldots,3$.
  Therefore, the algorithm returns the following set of states:
  \[
  \begin{array}{lll}
  \cS = \{ & \tuple{\pr{main Len xs},\nil},\;
           \tuple{\pr{lenmax xs},\;[(\pr{fst x},\pr{x})]}, \\
          &  \tuple{\pr{fst (len xs, max xs)},\nil},\;
           \tuple{\pr{len xs},\nil} & \}
  \end{array}
  \]
\end{example}
The total correctness of the algorithm in Fig.~\ref{pe-alg2} is stated
in the following theorem:

\begin{theorem} \label{termination}
  Given a flat program $\cR$ and an initial term $t$, the algorithm in
  Fig.~\ref{pe-alg2} terminates computing a set of states $\cS$ such
  that $\tuple{t,\nil}$ is $\cS$-closed.
\end{theorem}

\section{Extraction of the Slice} \label{extract}

In this section, we introduce the final step of our slicing process,
i.e., the extraction of the program slice. Let us recall that it must
be a \emph{fragment} of the original program|thus no instantiation of
variables is allowed|and produce the same outputs for the slicing
criterion as the original program. Here, we follow the
\emph{simplified} form for program slices, i.e., case branches of the
form $p \to \top$ are deleted, and function definitions of the form
$f(\ol{x_n}) = \top$ do not appear in the slice.

First, we need the following auxiliary function that returns the terms
which are relevant in order to extract a program slice from the set of
states computed by the algorithm of Fig.~\ref{pe-alg2}:

\begin{definition}[residual calls]
  Let $\cS$ be a set of states returned by the algorithm of
  Fig.~\ref{pe-alg2} and let
  $
  T_{\cS} = \{ t \mid \tuple{t,S} \in \cS \}
  $.
  Then, the set of residual calls of $\cS$ is defined as follows:
  \[
  \mbox{}\hspace{-3ex}\foo{residual\_calls}(\cS) = T_{\cS} \cup \{ t' \mid
  \tuple{t,S} \in \cS,~t' \in \foo{calls}(S),~\mbox{\sf and $t'$ is
    not $T_{\cS}$-closed} \}
  \]
\end{definition}
Observe that, in the above definition, $\foo{residual\_calls}$ should
also return the function calls in the computed stacks when they are
not closed w.r.t.\ the set of first components of the states in $\cS$.
This is mandatory in order to ensure a full equivalence w.r.t.\ the
standard semantics.
Program slices can now be built as follows:

\begin{definition}[construction of program slices] \label{cons-slice}
  Let $\cS$ be a set of states returned by the algorithm of
  Fig.~\ref{pe-alg2}. Then, a program slice is obtained from
  $
  \foo{build\_slice}(\foo{residual\_calls}(\cS))
  $,
  where function $\foo{build\_slice}$ is defined as follows:
  \[
  \begin{array}{lll}
    \foo{build\_slice}(\{\:\}) & = & \{\:\} \\
    \foo{build\_slice}(\{f(\ol{t_{n}})\} \cup T') & = &
          \{f(\ol{x_{n}}) = e'\} \cup \foo{build\_slice}(T') \\
  & & \mbox{\sf where } f(\ol{x_{n}}) = e \in \cR,\; \rho=\{\ol{x_{n} \mapsto t_{n}}\},\\
  & & \mbox{\sf and } 
  \sql e\sqr \rho \longrightarrow^\ast e' \not\longrightarrow
  \end{array}
  \]
\end{definition}
The new calculus which is used to construct the rules of the slice is
depicted in Fig.~\ref{simplify-rules}.  First, note that the symbols
``$\sql$'' and ``$\sqr$'' in an expression like $\sql e \sqr \rho$ are
purely syntactical, i.e., they are only used to mark subexpressions
where the inference rules may be applied. The substitution $\rho$ is
used to store the bindings for the program variables. Let us briefly
explain the rules of the new calculus.

Rule {\sf var} simply returns a variable unchanged. Rule {\sf cons}
applies to constructor-rooted terms; it leaves the outermost
constructor symbol and (recursively) inspects the arguments.

Rules {\sf select} and {\sf guess} proceed similarly to their
counterpart in Fig.~\ref{smallstep} but leave the case structure
untouched; the substitution $\rho$ is used to check the current value
of the case argument.  We only deal with \emph{variable} case
arguments since the considered expression is the right-hand side of
some program rule (cf.\ Fig.~\ref{flat}). Note that rule {\sf guess}
is now deterministic (and, thus, the entire calculus).

Finally, rules {\sf fun} and {\sf remove} are used to reduce function
calls: when there is some term in $\foo{residual\_calls}(\cS)$ with
the same outermost function symbol, 
we proceed as in rule {\sf cons}; otherwise, we return $\top$ (which
means that the evaluation of this function call is not needed).

\begin{figure}[t]
\begin{center}
$
\begin{array}{l@{~~~}l@{~\longrightarrow~}l}
\mbox{Rule} & \Control & \Control \\\hline\\[-2ex]
\mbox{\sf var} & \sql x \sqr \rho & x   \;\\
\mbox{\sf cons} & \sql c(\ol{t_{n}}) \sqr \rho  
& c(\sql t_{1} \sqr \rho, \ldots, \sql t_{n} \sqr \rho)   \;\\
\mbox{\sf select} & \sql (f)case\;x\;of\;\{\ol{p_{k} \to e_{k}}\} \sqr 
\rho & (f)case\;x\;of\;\{p_{i} \to \sql e_{i} \sqr \rho' \}   \;\\
\mbox{\sf guess} & \sql (f)case\;x\;of\;\{\ol{p_{k} \to e_{k}}\} \sqr 
\rho & (f)case\;x\;of\;\{\ol{p_{k} \to \sql e_{k} \sqr \rho_k }\}  \;\\
\mbox{\sf fun} & \sql f(\ol{t_{n}}) \sqr \rho
& f(\sql t_{1} \sqr \rho,\ldots,\sql t_{n} \sqr \rho)  \;\\
\mbox{\sf remove} & \sql f(\ol{t_{n}}) \sqr \rho
&  \top  \;\\\hline
\end{array}$

$\begin{array}{@{}l@{~}ll}
  \\[-2ex]
  \mbox{where in} &\mbox{{\sf select}: }& \rho(x) = c(\ol{t_{n}}),\;
  p_i = c(\ol{x_n}), \; \rho' = \{\ol{x_n \mapsto t_n}\} \circ \rho,
  \mbox{ and }
  i\in\{1,\ldots,k\} \\
  &\mbox{{\sf guess}: }& \rho(x) \in \cX,\; \rho_i = \{x \mapsto p_i\}\circ\rho, \mbox{ and } 
   i \in \{1,\ldots,k\} \\
  &\mbox{{\sf fun}: }& \mbox{there is some term in $\foo{residual\_calls}(\cS)$ rooted by $f$}\\
  &\mbox{{\sf remove}: }& \mbox{otherwise}
\end{array}
$
\end{center}
\vspace{-2ex}
\caption{Simplified Unfolding Rules}
\label{simplify-rules}
\vspace{-2ex}
\end{figure}

\begin{example} \label{lenapp3}
  Consider the set of states computed in Example~\ref{lenapp2}.  From
  this set, function \emph{residual\_calls} returns the set of terms:
  \[
  \{ \pr{main Len xs},\; \pr{lenmax xs},\; \pr{fst (len xs, max xs)},\;\pr{len xs} \}
  \]
  Now, we construct a residual rule for each term of the set. For
  instance, for the term ``\pr{main Len xs}'', the associated residual
  rule is:
  \[
  \pr{main op xs} \;=\; \pr{fcase op of } \{ \pr{Len} \to \pr{fst (lenmax xs)} \}
  \]
  since the following derivation can be performed (with $\rho = \{
  \pr{op} \mapsto \pr{Len} \}$):\\[2ex]
  $
  \begin{array}{l}
    \sql \pr{fcase op of } \{ \pr{ Len} \to \pr{fst (lenmax xs)};
                           \pr{ Max} \to \pr{snd (lenmax xs)} \;\} \sqr \rho\\
    \hspace{10ex}\longrightarrow_{\mathsf{select}} 
       \hspace{1ex}\pr{fcase op of } \{ \pr{ Len} \to \sql \pr{fst (lenmax xs)}\sqr \rho \;\} \\
    \hspace{10ex}\longrightarrow_{\mathsf{fun}} 
       \hspace{3ex}\pr{fcase op of } \{ \pr{ Len} \to \pr{fst } (\sql \pr{lenmax xs} \sqr \rho) \;\} \\
    \hspace{10ex}\longrightarrow_{\mathsf{fun}} 
       \hspace{3ex}\pr{fcase op of } \{ \pr{ Len} \to \pr{fst }  (\pr{lenmax } \sql \pr{xs} \sqr \rho) \;\} \\
    \hspace{10ex}\longrightarrow_{\mathsf{var}} 
       \hspace{3ex}\pr{fcase op of } \{ \pr{ Len} \to \pr{fst (lenmax xs)} \;\} \\
  \end{array}
  $\\[2ex]
  By constructing a residual rule associated to each of the remaining
  terms, the computed slice coincides with the (simplified version of
  the) program slice which is shown in Example~\ref{lenapp}.
\end{example}
Now, we show that the result of Definition~\ref{cons-slice} is a
program slice of the original program according to
Definition~\ref{prog-abs}.

\begin{theorem}  \label{abs-th}
  Let $\cR$ be a flat program and $t$ a term.  Let $\cS$ be a set of
  states computed by the algorithm of Fig.~\ref{pe-alg2} from $\cR$
  and $t$. Then, $\cR' =
  \foo{build\_slice}(\foo{residual\_calls}(\cS))$ is a program slice of
  $\cR$, i.e., $\cR' \succeq \cR$.
\end{theorem}
Finally, the correctness of the computed slices (according to
Def.~\ref{slice-def}) is inherited by the correctness of the
underlying partial evaluation process.

\begin{theorem} \label{final}
  Let $\cR$ be a flat program and $t$ a term.  Let $\cS$ be a set of
  states computed by the algorithm of Fig.~\ref{pe-alg2} from $\cR$
  and $t$. If computations for $t$ in $\cR$ do not suspend, then $t$
  computes the same values and answers in $\cR$ and in
  $\foo{build\_slice}(\foo{residual\_calls}(\cS))$.
\end{theorem}

\section{Implementation}
\label{exp}

\begin{table}
\caption{Partial evaluator vs program slicer --- code structure} \label{pe-slicer}
\begin{minipage}{\textwidth}
\begin{tabular}{lrrrrrrr} \hline\hline
& & \texttt{main} & \texttt{global} & \texttt{local} & \texttt{post} 
 & \texttt{util} & \textbf{Total} \\\hline
Partial evaluator  & \emph{(lines)} & 306 & 403 & 888 & 433 & 316 & \textbf{2346} \\
            & \emph{(functions)} &  22 &  43 &  83 &  44 & 38  & \textbf{230} \\[2ex]
Program slicer & \emph{(lines)} & 232 & 486 & 249 & 195 & 419 & \textbf{1581} \\ 
            & \emph{(functions)} &  20 &  50 &  29 &  26 &  55 & \textbf{180} \\ \hline\hline
\end{tabular}
\end{minipage}
\end{table}

In order to check the practicality of the ideas presented so far, a
prototype implementation of the program slicer for Curry programs has
been developed in Curry itself. The resulting tool covers not only the
flat programs of Sect.~\ref{sec-foundations} but also source Curry
programs (which are automatically translated to the flat syntax).
Moreover, it also accepts higher-order functions, overlapping
left-hand sides, several predefined (built-in) functions, etc. The
implemented tool is publicly available from {\tt
  http://www.dsic.upv.es/users/elp/german/slicing/}.

It is worthwhile to note that the development of the program slicer
required a small implementation effort since it was developed by
extending an existing partial evaluator for Curry programs
\cite{AHV02}.  Table~\ref{pe-slicer} shows the structure of both the
partial evaluator and the program slicer, including the lines of code
and the number of functions for each basic component:
\begin{description}
\item[\texttt{main:}] basic definitions and data type declarations,
  reading of source program, writing of transformed program, etc;
  
\item[\texttt{global:}] global control, including termination tests and
  generalization operations;
  
\item[\texttt{local:}] local control, i.e., a non-standard
  meta-interpreter;
  
\item[\texttt{post:}] post-processing transformation, i.e., renaming
  and post-unfolding compression in the partial evaluator and
  extraction of the slice in the program slicer;
  
\item[\texttt{util:}] general utilities and pretty printing.
\end{description}
Basically, components \texttt{main}, \texttt{local}, and \texttt{util}
were almost straightforwardly adapted from the partial evaluator to
the program slicer. For instance, component \texttt{local} of the
program slicer|which corresponds to the semantics shown in
Fig.~\ref{smallstep}|is a simplified version of the same component in
the partial evaluator, since only a one-step unfolding is required
here. More significant changes were made in component \texttt{global}.
In contrast to the partial evaluator, the program slicer introduces
the use of \emph{states} and, thus, it required the implementation of
rules \textsf{replace} and \textsf{flatten}, as well as the associated
abstraction operator.  Finally, component \texttt{post} of the partial
evaluator was entirely replaced, since the program slicer does not
perform neither renaming nor post-unfolding compression but should
only extract the residual rules according to the calculus of
Fig.~\ref{simplify-rules}.

Our slicing tool is able to compute the slice of Example~\ref{lenapp},
thus it is strictly more powerful than naive approaches based on
graphs of functional dependences. In general, forward slicing has been
proved particularly useful in the areas of program understanding, dead
code removal, and code reuse. Now, we illustrate the application of
the program slicer with some selected examples.
First, we consider the program of Example~\ref{lenapp} (in Curry
syntax):
\startprog
main Len xs = fst (lenmax xs)
main Max xs = snd (lenmax xs)\\[-1ex]

lenmax xs = (len xs, max xs)\\[-1ex]

len []     = Z
len (x:xs) = Succ (len xs)\\[-1ex]

max [x]      = x
max (x:y:ys) = if (x $\sleq$ y) then max (y:ys)
                           else max (x:ys)

Z $\sleq$ m               = True
(Succ n) $\sleq$ Z        = False
(Succ n) $\sleq$ (Succ m) = n $\sleq$ m\\[-1ex]

fst (a,b) = a
snd (a,b) = b
\stopprog
Given the slicing criterion ``\pr{main Len xs}'', our tool returns the
following slice:
\startprog
main Len xs = fst (lenmax xs)\\[-1ex]

lenmax xs = (len xs, $\top$)\\[-1ex]

len []     = Z
len (x:xs) = Succ (len xs)\\[-1ex]

fst (a,b) = a
\stopprog
Here, the second rule of function \pr{main} as well as the definitions
of functions \pr{max}, \pr{$\sleq$}, and \pr{snd} have been sliced
away, since they are not needed when the first parameter of \pr{main}
is the constant \pr{Len}. Note that the removal of \emph{case
  branches} in the flat language is now viewed in Curry as the removal
of \emph{rules} in a function definition, e.g.,
\startprog
main op xs = fcase op of $\{$ Len $\to$ fst (lenmax xs);
                           Max $\to$ $\top$ $\}$
\stopprog
is simply written as follows:
\startprog
main Len xs = fst (lenmax xs)
\stopprog
Let us now consider a similar situation but in a higher-order context:
\startprog
trans p xs = map (f p) xs\\[-1ex]

map f []     = []
map f (x:xs) = f x : map f xs\\[-1ex]

f A = inc
f B = dec
f C = square
...\\[-1ex]

inc x = Succ x
dec (Succ x) = x
square x = x * x
...
\stopprog
Function \pr{trans} applies a parametric function, \pr{f}, to all the
elements of a given list. Now, the computed slice w.r.t.\ the slicing
criterion ``$\tt trans\;A\;xs$'' is as follows:
\startprog
trans p xs = map (f p) xs\\[-1ex]

map f []     = []
map f (x:xs) = f x : map f xs\\[-1ex]

f A = inc\\[-1ex]

inc x = Succ x
\stopprog
Again, all functions but \pr{inc} and the first rule for \pr{f} have
been deleted, which shows that our approach works well in the presence
of higher-order functions.
Finally, let us show an example which illustrates the removal of dead
code due to lazy evaluation. Consider the following program:
\startprog 
lenInc n xs = len (incL n xs)\\[-1ex]

len []     = Z
len (x:xs) = Succ (len xs)\\[-1ex]

incL n []     = []
incL n (x:xs) = inc n : incL n xs\\[-1ex]

inc x = Succ x
\stopprog
Here, function \pr{lenInc} takes a number and a list, and returns the
length of the list which results from adding the given number to each
element of the original list. Clearly, in a lazy context, function
\pr{inc} will never be executed. Therefore, the computed slice w.r.t.\ 
``$\tt lenInc\;n\;xs$'' (i.e., no input data is known) is as follows:
\startprog 
lenInc n xs = len (incL n xs)\\[-1ex]

len []     = Z
len (x:xs) = Succ (len xs)\\[-1ex]

incL n []     = []
incL n (x:xs) = $\top$ : incL n xs\\[-1ex]
\stopprog
The occurrence of $\top$ in the definition of \pr{incL} shows that the
values of the elements in the list are not needed to compute the
length of the given list.

\begin{table} 
\caption{Partial evaluator vs program slicer --- selected benchmarks} \label{times}
\begin{minipage}{\textwidth} \footnotesize
\begin{tabular}{lrrrrrrrr} \hline\hline
& \multicolumn{1}{r}{\textsf{PE}} & \multicolumn{1}{r}{\textsf{Slicing}} 
 & \multicolumn{3}{c}{\textsf{Runtime}} & \multicolumn{3}{c}{\textsf{Size}} \\ \cline{4-6} \cline{7-9}
 & \multicolumn{1}{r}{\textsf{time}} 
  & \multicolumn{1}{r}{\textsf{time}} 
 & \textsf{Orig} & \textsf{Spec} & \textsf{Sliced} & \textsf{Orig} 
 & \textsf{Spec} & \textsf{Sliced} \\ 
\textsf{Benchmark} & \textsf{ms} & \textsf{ms}   & \textsf{ms} & \textsf{\%}
   & \textsf{\%} & \textsf{bytes} & \textsf{\%} & \textsf{\%} \\ \hline
\texttt{ackermann} & 20490 & 1370 & 1330 & 98.50\% & 99.25\% & 2039 & 228.3\% & 49.93\% \\
\texttt{allones} & 5090 & 1990 & 1140 & 111.40\% & 100.88\% & 3502 & 165.25\% & 63.42\% \\
\texttt{fibonacci} & 150 & 270 & 380 & 81.58\% & 102.63\% & 2438 & 64.93 & 50.82\% \\
\texttt{filtermap} & 280 & 450 & 1460 & 84.93\% & 100.00\% & 2147 & 28.50\% & 69.26\% \\
\texttt{fliptree} & 2800 & 1230 & 1430 & 92.31\% & 93.71\% & 2619 & 202.33\% & 52.20\% \\
\texttt{foldr.map} & 80 & 320 & 630 & 60.32\% & 100.00\% & 1784 & 21.41\% & 64.63\% \\
\texttt{foldr.sq} & 70 & 310 & 670 & 65.67\% & 101.49\% & 1763 & 21.10\% & 64.61\% \\
\texttt{foldr.sum} & 6730 & 1400 & 1570 & 80.25\% & 98.09\% & 4678 & 35.85\% & 13.92\% \\
\texttt{funinter} & 504930 & 2220 & --- & --- & --- & 5288 & 657.19\% & 61.86\% \\
\texttt{gauss} & 11680 & 950 & 700 & 82.86\% & 98.57\% & 2115 & 61.56\% & 42.36\% \\
\texttt{iterate} & 1950 & 750 & 890 & 25.84\% & 103.37\% & 1968 & 117.99\% & 69.61\% \\
\texttt{kmpAAB} & 710 & 990 & 380 & 31.58\% & 105.26\% & 3348 & 42.29\% & 75.30\% \\
\texttt{kmpAAAAAAB} & 9870 & 3250 & 790 & 35.44\% & 100.00\% & 3968 & 104.86\% & 69.78\% \\
\texttt{power} & 12710 & 890 & 620 & 95.16\% & 103.23\% & 2830 & 203.29\% & 46.93\% \\
\texttt{quicksort}& 450 & 670 & 260 & 165.38\% & 103.85\% & 2711 & 84.06\% & 81.48\% \\
\texttt{reverse}& 4590 & 970 & 680 & 98.53\% & 101.47\% & 1873 & 251.09\% & 53.87\% \\
\hline
\textbf{Average} & \textbf{36411} & \textbf{1127} 
  & \textbf{862} & \hspace{-2ex}\textbf{80.65\%} & \hspace{-2ex}\textbf{100.79\%} 
  & \textbf{2817} & \hspace{-2ex}\textbf{143.13\%} & \hspace{-2ex}\textbf{58.12\%} \\
\hline\hline
\end{tabular}
\end{minipage}
\vspace{-2ex}
\end{table}

Let us mention that, in contrast to the original partial evaluator,
the implemented program slicer can deal with larger programs
efficiently. This is mainly due to the monovariant/monogenetic nature
of the underlying partial evaluator, which simplifies the computation
of a closed set of terms. Table~\ref{times} shows a summary of the
experiments conducted on an extensive set of benchmarks. We used the
Curry$\to$Prolog compiler of PAKCS 1.6.0 \cite{PAKCS00} running on a
2.4 GHz Linux-PC (Intel Pentium IV with 512 KB cache). Runtime input
goals were chosen to give a reasonably long overall time. Code size
was obtained by measuring the intermediate FlatCurry files (suffix
\texttt{.fcy}) generated by PAKCS. The considered benchmarks are
available from {\tt http://www.dsic.upv.es/users/elp/german/slicing/}.

The results in Table~\ref{times} show that the program slicer is in
almost all cases much faster than the partial evaluation tool. As
expected, the runtime of the sliced programs do not significantly
differ from the runtimes of the original ones, since only some program
rules (or expressions) have been deleted; this shows that little
overhead has to be paid for adding extra functions to a program.
Anyway, the main purpose of slicing is not speedup, but reducing code
size.  In this case, slicing has managed an overall code size
reduction of 57.60\% whereas the partial evaluator has increased the
code size by 162.26\%. Indeed, the slicing never increases the code
size, while the partial evaluator has increased the code size by
657.18\% in the worst case. On the other hand, there are cases where
specialization achieves much smaller code size than slicing, e.g., for
\texttt{filtermap} where the specializer has managed to transform the
composition of several higher-order functions into a single
first-order function.

\section{Related Work}
\label{related}

Although program slicing was originally introduced in the imperative
programming setting, it has been applied to almost all programming
paradigms, e.g., object-oriented programs \cite{TCFR96,LH96,Ste98},
logic programs \cite{SD96,ZCU01}, functional programs
\cite{Bis97,FT98}, or algebraic specifications \cite{WA98}. Although
we are not aware of any previous work addressing forward slicing of
multi-paradigm functional logic programs, in the following we review
the closest approaches to our work.

Within imperative programming, the closest approach is that of Blazy
and Facon \citeyear{BF98}, who use partial evaluation for program
understanding in Fortran. Since they do not want to change the
original structure of the code, no unfolding is performed (similarly
to our one-step unfoldings). Also, they neither introduce new
variables nor rename the existing ones. In this work, we have followed
a very similar approach in order to define a forward slicing algorithm
for functional logic programs. In both approaches, a simplified
partial evaluator that does not change the structure of the original
program has been introduced.

Within the logic programming paradigm, Gyim\'othy and Paakki
\citeyear{GP95} introduce the first approach to slicing. They define a
specific slicing algorithm which computes a slice of the proof tree in
order to reduce the number of questions asked by an algorithmic
debugger \cite{Sha83}. The slice is computed from a static dependency
graph containing only oriented data dependencies. In contrast to our
work, their algorithm cannot be not used to compute \emph{executable}
programs.  Schoening and Ducass\'e \citeyear{SD96} define the first
(backward) slicing algorithm for Prolog programs which produce
executable slices. They introduce an abstraction relation in order to
formalize the notion of program slice. Our notion of slice in
Section~\ref{pe} is somehow inspired by this work. Leuschel and
S{\o}rensen \citeyear{LS96} introduce the concept of \emph{correct
  erasure} in order to detect and remove redundant arguments from
logic programs. They present a constructive algorithm for computing
correct erasures which can be used to perform a simple form of
slicing. Actually, Leuschel and Vidal \citeyear{LV05} have very
recently introduced a new approach to forward slicing of logic
programs which is based on a combination of the ideas presented in
this work and the redundant argument filtering of Leuschel and
S{\o}rensen \citeyear{LS96}.

As for functional programs, Field and Tip \citeyear{FT98} present a
very detailed study of the concept of slicing associated with
left-linear term rewriting systems (a notion of ``program'' very close
to the one considered in our work). Their definition of slice is also
based on a notion of \emph{neededness} but, in contrast to our work,
they consider {backward} slicing (and compute slices that are not
executable on the standard interpreter). Another closely related
approach has been introduced by Reps and Turnidge \citeyear{RT96}.
They define a {backward} slicing technique for functional programs
which can be used to perform a sort of program specialization that
cannot be achieved by standard partial evaluation. Their work can be
seen as complementary to ours, since we are interested in the use of
partial evaluation to perform program slicing. On the other hand,
Hallgren \citeyear{Hal03} reports some experiments with a Haskell
slicer. It is mainly based on the construction of a graph of
functional dependences and, thus, it is less powerful than our partial
evaluation-based slicer.

Very recently, Ochoa et al.\ \citeyear{OSV04} have introduced a novel
approach to \emph{dynamic} backward slicing of functional logic
programs which is based on an extension of the tracing technique of
Bra{\ss}el et al.\ \citeyear{BHHV04}. In particular, their approach
relies on constructing a \emph{redex trail} of a given computation in
order to compute all program dependences. Basically, a redex trail is
a directed graph which records copies of all values and redexes of a
computation, with a backward link from each reduct to the parent redex
that created it.  Then, a backward slice can easily be obtained by
mapping the relevant nodes of the redex trail to concrete locations of
the source program.  This approach has also been applied to Haskell
programs by Chitil \citeyear{Chi04}. These approaches are not based on
partial evaluation but on well-known techniques for debugging
functional programs.  Therefore, the implementation of a
\emph{dynamic} slicer is relatively easy if one already has a debugger
based on redex trails. However, they are not useful in order to
develop a \emph{static} slicing tool. In contrast, our approach can be
used to perform both static and (forward) dynamic slicing.

\section{Conclusions and Future Work}
\label{sec-concl}

This work introduced the first approach to forward slicing of
multi-paradigm (functional logic) programs. Although some extensions
were needed, our developments basically rely on adapting and extending
an online partial evaluation scheme for such programs. Thus, the
implementation of an associated slicing tool was easily achieved by
extending an existing partial evaluation tool. Moreover, our approach
helps to clarify the relation between program slicing and partial
evaluation in a functional logic context.  The application of our
developments to (first-order) lazy functional programs would be
straightforward, since the considered language is a conservative
extension of a pure lazy functional language and the (online) partial
evaluation techniques are similar, e.g., positive supercompilation
\cite{SGJ93}. On the other hand, similar ideas have already been
applied to define a forward slicing technique for logic programs
\cite{LV05}.

An interesting topic for future work is the extension of our approach
to perform \emph{backward} slicing. Here, the computed slice should
contain those program statements which are needed to compute some
selected \emph{fragment} of the output. While forward slicing is
useful for program understanding, reuse, maintenance, etc., backward
slicing can be applied to, e.g., program debugging, specialization and
merging.

\subsection*{Acknowledgements}

We gratefully acknowledge the anonymous referees as well as the
participants of LOPSTR 2002 for many useful comments and suggestions.
We also thank Michael Leuschel for his helpful remarks concerning the
relation between partial deduction and slicing in the context of logic
programming.




\newpage
\appendix

\section{Proofs of technical results} \label{proofs}

\mbox{}\\
\textsl{Lemma \ref{flattened-states}}\\
Let $s$ be a flattened state. Then $s$ has the form 
$\tuple{v,\nil}$, where $v$ is a value, or $\tuple{f(\ol{t_{n}}),S}$, 
where $f(\ol{t_{n}})$ is an operation-rooted term.

\begin{proof}
  We prove the claim by contradiction. Let $s$ be a flattened state of
  the form $\tuple{v,S}$ where $v$ is a value and $S$ is not empty.
  Then rule {\sf replace} could be applied to $s$, thus contradicting
  the hypothesis of the lemma. Thus, $s$ should be of the form
  $\tuple{v,\nil}$ or $\tuple{e,S}$, where $e$ is not a value. To show
  that $e$ must be an operation-rooted term, it suffices to consider
  that rules {\sf replace} and {\sf flatten} do not return case
  expressions (only operation-rooted terms) and that the initial state
  cannot contain case expressions (since it was returned by the
  operator $\foo{unfold}$).
\end{proof}

\mbox{}\\
\textsl{Lemma \ref{safe-abstract}}\\
Let $\cS$ be a set of flattened states and $\cS'$ a set of unfolded 
states (as returned by unfold). Then the 
states in $\cS \cup \cS'$ are closed w.r.t.\ 
$\foo{abstract}(\cS,\cS')$.\\

\noindent
In order to prove this lemma, we first need the following preparatory
definitions and results.  We use the notation $depth(t)$ to denote the
maximum number of nested symbols in the term $t$. Formally, if $t$ is
a constant or a variable, then $depth(t) = 1$. Otherwise,
$depth(f(\ol{t_{n}})) = 1 + max (\{depth(t_1),\ldots, depth(t_n)\})$.
The following result establishes the transitivity of the closedness
relation on terms.

\begin{proposition}[Alpuente et al.\ 1998] \label{transi-closs} 
  If term $t$ is $T_1$-closed, and the terms in $T_1$ are
  $T_2$-closed, then $t$ is $T_2$-closed.
\end{proposition}
We define the complexity $\cM_{T}$ of a set of terms $T$ as the finite 
multiset of natural numbers corresponding to the depth of the
elements of $T$. Formally, $\cM_{T}= \{depth(t)\mid t\in T\}$.
We consider the well-founded total ordering $\mathop{<}_{mul}$ over
multiset complexities by extending the well-founded ordering $<$ on 
$\nat$ to the set $M(\nat)$  of finite multisets over $\nat$. The 
set $M(\nat)$ is well-founded under the ordering $\mathop{<}_{mul}$ 
since $\nat$ is well-founded under $<$.
Let $\cM, \cM'$ be multiset complexities, then: $\cM \mathop{<}_{mul}
\cM' \;\Leftrightarrow\; \exists X \subseteq \cM, X' \subseteq \cM'$
such that $\cM = (\cM' - X') \cup X\:$ and $\:\forall n \in X,
\:\:\exists n' \in X'$ such that $n < n'$. This ordering is naturally
extended to sets of states by simply considering the terms represented
by the states in each set.

Now, we can proceed with proof of Lemma~\ref{safe-abstract}. We follow
the scheme of the proof of Lemma~5.13 in \cite{AFV98} but extend it to
deal with states:

\begin{proof}
We proceed by structural induction on $\cS \cup \cS'$. Since the base case is 
trivial ($\cS$ is always $\cS$-closed), we consider the inductive 
case. Let $\cS'' = \{s'_{1},\ldots,s'_{n}\}$, $n>1$, be the set of 
states resulting from flattening the states in $\cS'$, i.e., 
$\cS'' = \{ s'_{i} \mid s_{i} \in \cS' \mbox{ and } s_{i} 
\Longrightarrow^\ast_{\sf replace/flatten} 
s'_{i} \not\Longrightarrow_{\sf replace/flatten} \}$. Trivially, we
have that $\cM_{\cS'} = \cM_{\cS''}$ and that $\cS \cup \cS'$ is
closed w.r.t.\ $\cS \cup \cS''$, since the process of flattening does
not change the terms represented by the states.  By the definition of
$\foo{abstract}$, we have the following equalities:
\[
\begin{array}{lll}
\foo{abstract}(\cS,\cS') & = & \foo{abstract}(\cS,\cS'') \\
& = & \foo{abs}(\foo{abs}(\ldots \foo{abs}(\cS,s'_{1})\ldots, 
  s'_{n-1}),s'_{n}) \\
& = & \foo{abs}(\foo{abstract}(\cS,\cS'' \:\backslash \:
\{s'_{n}\}), s'_{n}) \\
& = & \foo{abs}(\cS^\ast,s'_{n})
\end{array}
\]
where $\cS^\ast = \foo{abstract}(\cS,\cS'' \:\backslash \:
\{s'_{n}\})$ and
$s'_{n}$ is an arbitrary state of $\cS''$. By the 
inductive hypothesis, we know that $\cS \cup (\cS'' \:\backslash\: 
\{s'_{n}\})$ is closed w.r.t.\ $\cS^\ast$. Now, we proceed with the 
call $\foo{abs}(\cS^\ast,s'_{n})$. Here, we distinguish the 
following cases depending on the structure of $s'_{n}$:
\begin{description}
\item[\sf $s'_{n} = \tuple{x,\nil}$:] Then
  $\foo{abs}(\cS^\ast,s'_{n}) = \cS^\ast$ and the claim follows by
  Lemma~\ref{transi-closs}.

\item[\sf $s'_{n} = \tuple{c(\ol{t_{n}}), \nil}$:] Assume that
  $\ol{t'_{m}}$ are the maximal operation-rooted subterms of
  $c(\ol{t_{n}})$. Then $\foo{abs}(\cS^\ast,s'_{n}) =
  \foo{abstract}(\cS^\ast, \cS^c)$, with $\cS^c = \{
  \ol{\tuple{t'_{m},\nil}}\}$. Since $\cM_{\cS^\ast \cup \cS^c}$
  $\mathop{<}_{mul} \cM_{\cS^\ast \cup \{s'_{n}\}}$, the proof follows
  by Lemma~\ref{transi-closs} and the inductive hypothesis.
  
\item[\sf $s'_{n} = \tuple{f(\ol{t_{n}}),S}$:] Then, following the
  definition of function $\foo{abs}$, we consider three
  possibilities:
\begin{itemize}
\item If there is no state in $\cS^\ast$ whose first component is 
rooted by $f$, then $\foo{abs}(\cS^\ast,s'_{n}) = \cS^\ast \cup 
\{s'_{n}\}$. Thus, the claim follows by Lemma~\ref{transi-closs}.

\item If the state is ignored (because it is already closed and it is not 
equal to any existing state), then $\foo{abs}(\cS^\ast,s'_{n}) = 
\cS^\ast$. 
Again, the claim follows trivially by Lemma~\ref{transi-closs}.

\item Otherwise, there exists some state $\tuple{f(\ol{t'_{n}}),S'}$ 
and\\[2ex]
$ 
\mbox{}\hspace{2em}
\foo{abs}(\cS^\ast,s'_{n}) = \foo{abstract}((\cS^\ast \: 
\backslash 
\:\{\tuple{f(\ol{t'_{n}}),S'}\})\cup\{\tuple{f(\ol{t''_{n}}),S'}\},\cS^f)
$\\[2ex] 
where 
$\foo{msg}(\tuple{f(\ol{t'_{n}}),S'},\tuple{f(\ol{t_{n}}),S}) = 
(\tuple{f(\ol{t''_{n}}),S'},\cS^f)$,
$\cS^f = \foo{calls}(\sigma_{1}) \cup$ \linebreak
$\foo{calls}(\sigma_{2}) \cup \foo{calls}(S)$,
$\foo{msg}(f(\ol{t'_{n}}), f(\ol{t_{n}})) = f(\ol{t''_{n}})$,
$\sigma_{1}(f(\ol{t''_{n}})) = f(\ol{t'_{n}})$, and
$\sigma_{2}(f(\ol{t''_{n}})) = f(\ol{t_{n}})$. Now, by definition of
function \emph{msg}, it is easy to check that $\cS^\ast \cup
\{s'_{n}\}$ is closed w.r.t.\ $(\cS^\ast \: \backslash
\:\{\tuple{f(\ol{t'_{n}}),S'}\})\cup
\{\tuple{f(\ol{t''_{n}}),S'}\}\cup \cS^f$ and that $\cM_{(\cS^\ast \:
  \backslash \:\{\tuple{f(\ol{t'_{n}}),S'}\})\cup
  \{\tuple{f(\ol{t''_{n}}),S'}\}\cup \cS^f} \mathop{<}_{mul}
\cM_{\cS^\ast \cup \{s'_{n}\}} $.  Therefore, the proof follows by
Lemma~\ref{transi-closs} and the inductive hypothesis.
\end{itemize}
\end{description}
\end{proof}

\mbox{}\\
\textsl{Theorem \ref{termination}}\\
  Given a flat program $\cR$ and an initial term $t$, the algorithm in
  Fig.~\ref{pe-alg2} terminates computing a set of states $\cS$ such
  that $\tuple{t,\nil}$ is $\cS$-closed.

\begin{proof}
  The $\cS$-closedness of $\tuple{t,\nil}$ is a direct consequence of
  Lemma~\ref{safe-abstract} and Proposition~\ref{transi-closs}.
  Lemma~\ref{safe-abstract} ensures that the arguments of the operator
  $\foo{abstract}$ are always closed w.r.t.\ the new set of states,
  while Proposition~\ref{transi-closs} guarantees that the closedness
  of the initial state is correctly propagated through the whole
  process.
  
  The termination of the algorithm can be derived from the following
  facts:
\begin{enumerate}
\item Each iteration of the algorithm is finite. The finiteness of the
  application of the unfolding operator is obvious (since only one
  function unfolding is allowed). The termination of one application
  of operator $\foo{abstract}$ can easily be proved by following
  the scheme of the proof of Lemma~\ref{safe-abstract}: the
  computation terminates since each recursive call to
  $\foo{abstract}$ uses a set of states which is strictly lesser
  than the previous call (w.r.t.\ $\mathop{<}_{mul}$).

\item The termination of the whole iterative process is a 
consequence of the following facts:
\begin{itemize}
\item The number of states in the current set $\cS_{i}$ cannot be
  greater than the number of different functions in the original
  program, since the abstraction operator ensures that there is only
  one state for each function symbol of the program. Thus, it cannot
  grow infinitely.
  
\item The number of states in the set $\cS_{i+1}$ is always equal to
  or greater than the number of states in the set $\cS_{i}$. This
  property is immediate, since we only remove states in the last case
  of the definition of $\foo{abstract}$ and, there, we replace one
  state by a new (generalized) state.
  
\item Finally, each time one state is replaced by a new one, the new
  state is equal to or smaller than the previous state (according to
  the ordering based on the depth of the terms). If the new state has
  the same depth than the old one, then the process would terminate,
  since they would be equal (modulo renaming). If it is strictly
  smaller, then it will eventually reach an state whose first
  component is of the form $f(\ol{x_{n}})$ and, hence, it cannot be
  generalized again.
\end{itemize}
\end{enumerate}
\end{proof}

\mbox{}\\
\textsl{Theorem \ref{abs-th}}\\
  Let $\cR$ be a flat program and $t$ a term.  Let $\cS$ be a set of
  states computed by the algorithm of Fig.~\ref{pe-alg2} from $\cR$
  and $t$. Then, $\cR' =
  \foo{build\_slice}(\foo{residual\_calls}(\cS))$ is a program slice of
  $\cR$, i.e., $\cR' \succeq \cR$.

\begin{proof}
  This result is an easy consequence of Def.~\ref{cons-slice} and the
  calculus in Fig.~\ref{simplify-rules}:
  \begin{itemize}
  \item If there is a rule $f(\ol{x_n}) = e \in \cR$ and there is no
    term $f(\ol{t_n}) \in \foo{residual\_calls}(\cS)$, then $\cR'$
    does not contain a definition for $f$, i.e., $f(\ol{x_n}) = \top
    \in \cR'$ and, trivially, $\top \succeq e$.
    
  \item Otherwise, $f(\ol{x_n}) = e \in \cR$ and $f(\ol{x_n}) = e' \in
    \cR'$ with $\sql e\sqr \rho \longrightarrow^\ast e'
    \not\longrightarrow$ and $\rho=\{\ol{x_{n} \mapsto t_{n}}\}$.
    Now, we prove that $e' \succeq e$ by induction on the length $l$
    of the derivation $\sql e\sqr \rho \longrightarrow^\ast e'$:
    \begin{description}
    \item[\sf Base case ($l = 1$).] In this case, we only have the
      following possibilities:
      \begin{itemize}
      \item $e$ is a variable; thus, $e' = e$ and the claim follows
        trivially.
      \item $e=c()$ is a constructor constant; then, $e' = e$ and the
        claim follows.
      \item $e = g(\ol{t'_m})$ is operation-rooted and there is no
        term in $\foo{residual\_calls}(\cS)$ rooted by $g$. Then, $e'
        = \top$ and $e' \succeq e$.
      \end{itemize}
    \item[\sf Induction case ($l > 1$).] Here, we distinguish the
      following cases:
      \begin{itemize}
      \item $e = c(t'_1,\ldots,t'_m)$ with $c \in \cC$ and $m>0$.
        Then, $e' = c(\sql t'_1\sqr\rho,\ldots,\sql t'_m\sqr\rho)$ and
        the claim follows by induction.
        
      \item $e = \foo{(f)case}\;x\;\foo{of}\;\{p_1\to e_1; \ldots;p_k
        \to e_k\}$ with $\rho(x) = c(\ol{t'_{m}})$ and $p_i =
        c(\ol{y_m})$. Then, $e' = \foo{(f)case}\;x\;\foo{of}\;\{p_1\to
        \top; \ldots;p_i \to \sql e_i\sqr \rho';\ldots;p_k \to \top\}$
        with $\rho' = \{\ol{y_m \mapsto t'_m}\} \circ \rho$. By the
        inductive hypothesis, we have $\sql e_i\sqr \rho' \succeq e_i$
        and, thus, $e' \succeq e$.

      \item $e = \foo{(f)case}\;x\;\foo{of}\;\{p_1\to e_1; \ldots;p_k
        \to e_k\}$ with $\rho(x) \in \cX$. Then, $e' =
        \foo{(f)case}\;x\;\foo{of}\;\{p_1\to \sql e_1 \sqr \rho_1;
        \ldots;p_k \to \sql e_k \sqr \rho_k\}$, with $\rho_i = \{x
        \mapsto p_i\}\circ\rho$, $i \in \{1,\ldots,k\}$. By the
        inductive hypothesis, we have $\sql e_i \sqr \rho_i \succeq
        e_i$ for all $i = 1,\ldots,k$. Therefore, $e' \succeq e$.
        
      \item $e = g(t'_1,\ldots,t'_m)$ with $g \in \cF$.  Then, either
        $e' = g(\sql t'_1\sqr\rho,\ldots,\sql t'_m\sqr\rho)$ and the
        claim follows by induction, or there is no term in
        $\foo{residual\_calls}(\cS)$ rooted by $g$ and $e' = \top
        \succeq e$.
      \end{itemize}
    \end{description}
  \end{itemize}
\end{proof}

\mbox{}\\
\textsl{Theorem \ref{final}}\\
  Let $\cR$ be a flat program and $t$ a term.  Let $\cS$ be a set of
  states computed by the algorithm of Fig.~\ref{pe-alg2} from $\cR$
  and $t$. If computations for $t$ in $\cR$ do not suspend, then $t$
  computes the same values and answers in $\cR$ and in
  $\foo{build\_slice}(\foo{residual\_calls}(\cS))$.

\begin{proof}
  We present an sketch of the proof; the complete formalization is not
  difficult but would require the introduction of many notions and
  results from the original narrowing-driven specialization framework.
  Basically, the proof proceeds in a stepwise manner as follows:
\begin{itemize}
\item First, we consider the construction of a specialized program
  from the states in $\cS$ following the standard approach to
  narrowing-driven partial evaluation.  Formally, we construct a
  residual program $\cR'$ by producing a residual rule of the form
  \[
  \foo{ren}(S[t]) = \foo{ren}(S[t'])
  \]
  for each term in $\{ S[t] \mid \tuple{t,S} \in \cS\}$, where
  $\tuple{t,S} \Longrightarrow_{\sf fun} \tuple{t'',S}
  \Longrightarrow^\ast_{\sf select/guess} \tuple{t',S}
  \not\Longrightarrow_{\sf select/guess}$.  Function $\foo{ren}$ is
  applied to rename expressions so that the resulting residual program
  fulfills the syntax of flat programs (in spite of the form of
  $S[t]$).  Basically, $\foo{ren}(S[t])$ returns a term
  $f(\ol{x_{n}})$ where $f$ is a fresh function symbol and
  $\ol{x_{n}}$ are the different variables of $S[t]$.  By the
  correctness of narrowing-driven partial evaluation (since $t$ is
  $\cS$-closed by Theorem~\ref{termination}), we know that all the
  computations for $\tuple{t,\nil}$ in the original program can also
  be done for $\tuple{\foo{ren}(t),\nil}$ in the residual program
  constructed so far.

\item Then we consider a new program $\cR''$ which is obtained from
  $\cR'$ as follows: each rule in $\cR'$ of the form $\foo{ren}(S[t])
  = \foo{ren}(S[t'])$ is replaced by a new rule $\foo{ren}(t) =
  \foo{ren}(t')$. This replacement is safe in our context since the
  one-step unfolding does not affect to the current stack. A precise
  equivalence between the computations in $\cR'$ and $\cR''$ can
  easily be established under the extended operational semantics of
  Fig.~\ref{smallstep}. Note that renaming is still necessary to
  ensure that residual rules fulfill the syntax of Fig.~\ref{flat}.
  
\item Now, we define a new program $\cR'''$ which is obtained from
  $\cR''$ as follows: each rule $\foo{ren}(t) = \foo{ren}(t')$, with
  $t = f(\ol{t_{n}})$, is replaced by a new rule $f(\ol{x_{n}}) = e'$,
  where $f(\ol{x_{n}}) = e$ is a rule of the original program $\cR$,
  $\ol{x_{n}}$ are fresh variables, $\sql e \sqr \rho
  \longrightarrow^\ast e' \not\longrightarrow$, and $\rho =
  \{\ol{x_{n} \mapsto t_{n}}\}$. There is no need to apply a renaming
  of expressions in this case, since the program so constructed
  already fulfills the syntax of Fig.~\ref{flat} (cf.\ 
  Theorem~\ref{abs-th}). Now, each derivation for
  $\tuple{\foo{ren}(t),\nil}$ in $\cR''$ can also be done for
  $\tuple{t,\nil}$ in $\cR'''$ using the extended operational
  semantics of Fig.~\ref{smallstep}.  This is justified by the fact
  that the only difference between the rules of $\cR''$ and $\cR'''$
  is that bindings are applied to program expressions in $\cR''$ while
  they are represented implicitly in $\cR'''$ by means of case
  expressions in the right-hand sides of the program rules.
  
\item Finally, we extend $\cR'''$ by adding a residual rule of the
  form $f(\ol{x_{n}}) = e'$ for each call $f(\ol{t_{n}}) \in \{ t'
  \mid \tuple{t,S} \in \cS,~t' \in \foo{calls}(S),~\mbox{ and $t'$ is
    not $T_{\cS}$-closed} \}$, where $T_{\cS} = \{ t \mid \tuple{t,S}
  \in \cS \}$, $f(\ol{x_{n}}) = e$ is a rule of the original program
  $\cR$, $\ol{x_{n}}$ are fresh variables, $\sql e \sqr \rho
  \longrightarrow^\ast e' \not\longrightarrow$, and $\rho =
  \{\ol{x_{n} \mapsto t_{n}}\}$.  The extended program coincides with
  the result of $\foo{build\_slice}(\foo{residual\_calls}(\cS))$.  The
  claim follows by checking that each derivation for $\tuple{t,\nil}$
  in $\cR'''$ using the extended semantics can also be performed for
  $t$ in $\foo{build\_slice}(T)$ using the standard operational
  semantics. Intuitively, this equivalence holds because the only
  difference---we ignore here the suspension of flexible case
  expressions since we only consider computations which do not
  suspend---between the standard and the extended operational
  semantics is that the unfolding of some outer function call is
  (possibly) delayed until a complete one-step evaluation is possible.
  Therefore, the same computations can be proved with both calculi,
  except when there is some inner call which never reduces to a value
  (due to an infinite derivation). However, we ensure the equivalence
  even in this case by adding residual rules for the calls in the
  stack components which are not $T_{\cS}$-closed, i.e., for those
  calls which have some inner call with a non-terminating derivation.
\end{itemize}
\end{proof}

\end{document}